\documentclass[sigconf, nonacm]{acmart}
\usepackage{algorithm}
\usepackage{algorithmicx}
\usepackage{algpseudocode}
\usepackage{placeins}
\usepackage{graphicx}
\usepackage{float}
\usepackage{subcaption}
\usepackage{balance}
\usepackage{enumitem}
\newcommand\vldbdoi{XX.XX/XXX.XX}

\newcommand\vldbpagestyle{plain} 
\usepackage{listings}
\usepackage{xcolor}
\usepackage{placeins}  
\makeatletter
\def\@authorfont{\Large}
\def\@affiliationfont{\normalsize}
\makeatother

\lstdefinelanguage{json}{
  basicstyle=\ttfamily\small,
  numbers=none,
  stepnumber=1,
  showstringspaces=false,
  breaklines=true,
  frame=single,
  literate=
   *{0}{{{\color{black}0}}}{1}
    {1}{{{\color{black}1}}}{1}
    {2}{{{\color{black}2}}}{1}
    {3}{{{\color{black}3}}}{1}
    {4}{{{\color{black}4}}}{1}
    {5}{{{\color{black}5}}}{1}
    {6}{{{\color{black}6}}}{1}
    {7}{{{\color{black}7}}}{1}
    {8}{{{\color{black}8}}}{1}
    {9}{{{\color{black}9}}}{1}
    {:}{{{\color{black}{:}}}}{1}
    {,}{{{\color{black}{,}}}}{1}
    {\{}{{{\color{black}{\{}}}}{1}
    {\}}{{{\color{black}{\}}}}}{1}
    {[}{{{\color{black}{[}}}}{1}
    {]}{{{\color{black}{]}}}}{1}
}

\begin{document}
\title{Enhancing OLAP Resilience at LinkedIn}


\renewcommand{\thefootnote}{\fnsymbol{footnote}}
\setcounter{footnote}{1}

\author{Praveen Chaganlal}
\authornote{These authors contributed equally to this work and are listed alphabetically.}
\affiliation{\institution{LinkedIn}\city{Sunnyvale}\country{USA}}
\email{pchaganlal@linkedin.com}

\author{Jia Guo}
\authornotemark[2] 
\affiliation{\institution{LinkedIn}\city{Sunnyvale}\country{USA}}
\email{jiaguo@linkedin.com}

\author{Vivek Iyer Vaidyanathan Iyer}
\authornotemark[2]
\affiliation{\institution{LinkedIn}\city{Sunnyvale}\country{USA}}
\email{vvaidyanathan@linkedin.com}

\author{Dino Occhialini}
\authornotemark[2]
\affiliation{\institution{LinkedIn}\city{Sunnyvale}\country{USA}}
\email{docchialini@linkedin.com}

\author{Siddharth Teotia}
\affiliation{\institution{LinkedIn}\city{Sunnyvale}\country{USA}}
\email{steotia@linkedin.com}

\author{Sonam Mandal}
\affiliation{\institution{LinkedIn}\city{Sunnyvale}\country{USA}}
\email{somandal@linkedin.com}

\author{Subbu Subramaniam}
\affiliation{\institution{LinkedIn}\city{Sunnyvale}\country{USA}}
\email{mcvsubbu@gmail.com}

\author{Tianqi Li}
\affiliation{\institution{LinkedIn}\city{Sunnyvale}\country{USA}}
\email{tiali@linkedin.com}

\author{Xiaxuan Gao}
\affiliation{\institution{LinkedIn}\city{Sunnyvale}\country{USA}}
\email{xiaxgao@linkedin.com}

\author{Florence Zhang}
\affiliation{\institution{StarTree}\city{Mountain View}\country{USA}}
\email{florencezhsh@gmail.com}

\begin{abstract}

Apache Pinot is a real-time OLAP database, built and open-sourced by LinkedIn in 2015. It is a critical system in LinkedIn's online infrastructure stack, delivering interactive analytics at high read throughput and sub-second query latencies for several site-facing use cases such as Who-Viewed-My-Profile~\cite{linkedin_who_viewed}, Advertiser and Creator Analytics~\cite{linkedin_marketing_analytics}, Talent Insights~\cite{linkedin_talent_insights}, and Session Feed Item Selection~\cite{linkedin_recommended_content} while also powering internal business metrics dashboards. Maintaining strict SLAs at our scale is challenged by constant operational entropy. We present a holistic resiliency framework for Apache Pinot at LinkedIn that unifies three major failure vectors into a single, layered defense. 

\textbf{Workload Resilience:} We introduce Query Workload Isolation (QWI) to eliminate "noisy neighbor" interference via fine-grained resource budgeting and enforcement, delivering predictable tail latencies for multi-tenant mixed workloads with under 1\% overhead. 

\textbf{Structural Resilience:} We present Zone-Aware Replica Placement and Impact-Free Rebalancing, to provide high availability across physical fault domains within a data center and zero-downtime movement of data without degrading latency and availability SLOs. 

\textbf{Runtime Resilience:} We describe Adaptive Server Selection (ADSS), which dynamically routes queries around "fail-slow" nodes using performance signals, eliminating such resiliency issues by 90\%. 

Together, these mechanisms provide a production-proven blueprint for scaling mission-critical online analytics infrastructure under continuous hardware and workload churn.
\end{abstract}
\maketitle

\footnotetext{All code available at \url{https://github.com/apache/pinot}}


\setlength{\textfloatsep}{4pt plus 1pt minus 1pt}   
\setlength{\floatsep}{4pt plus 1pt minus 1pt}        
\setlength{\intextsep}{4pt plus 1pt minus 1pt}       
\setlength{\dbltextfloatsep}{4pt plus 1pt minus 1pt}
\setlength{\dblfloatsep}{4pt plus 1pt minus 1pt}

\setlength{\abovecaptionskip}{2pt}  
\setlength{\belowcaptionskip}{0pt}  

\lstset{aboveskip=4pt,belowskip=4pt}

\setlist{topsep=2pt, itemsep=1pt, parsep=0pt}

\pagestyle{\vldbpagestyle}


\section{Introduction}
Modern applications that enable decision making via business, operational, or other key metrics require both low query latencies and strict data freshness guarantees on several hundreds of TB to PB scale datasets with high query and ingestion throughput. To meet these requirements, OLAP database engines like Pinot, Druid, ClickHouse, and AnalyticDB that can deliver high speed analytics in real-time become foundational components in the infrastructure stack for an enterprise. For example, Alibaba's AnalyticDB deployment spans more than 2,000 nodes and serves 10+ PB of data. Pinot was built at LinkedIn to serve real-time analytics for heterogeneous online workloads with tight P99 SLAs. However, at this scale, the "steady state" is an illusion; maintaining these SLAs is a constant battle against operational entropy. We define this entropy as inevitable failures and performance fluctuations arising from shared-tenant resource contention, physical zone-level maintenance, and transient "fail-slow" hardware behavior. In this paper, we present a production-proven holistic resiliency framework for Apache Pinot that unifies these challenges into a layered architecture. Rather than treating resource management, data placement, and query routing as disjoint problems, we show how they form a cohesive system designed to absorb the constant churn of a petabyte-scale deployment.

\subsection{Apache Pinot Architecture Overview}
\label{sec:pinot-arch}

Apache Pinot~\cite{pinot2018} is a real-time distributed OLAP system designed for low-latency analytics. Pinot consumes events from streaming ingestion pipelines such as Kafka~\cite{kafka_website}, Kinesis~\cite{aws_kinesis}, and Pub/Sub~\cite{google_pubsub}, or from data lakes~\cite{DBLP:journals/corr/abs-2106-09592} via batch processing systems such as Spark~\cite{apache_spark}, Hive~\cite{apache_hive_docs}, and MapReduce~\cite{hadoop_mapreduce_tutorial}. Pinot indexes data as soon as it is ingested, providing a platform for building real-time dashboards on continuous streaming data. Pinot can achieve query latencies as low as 10ms at the 99th percentile.

The system is organized as three logical tiers (Figure~\ref{fig:pinot-arch}):

\begin{figure}[b]
  \centering
  \includegraphics[width=0.85\linewidth]{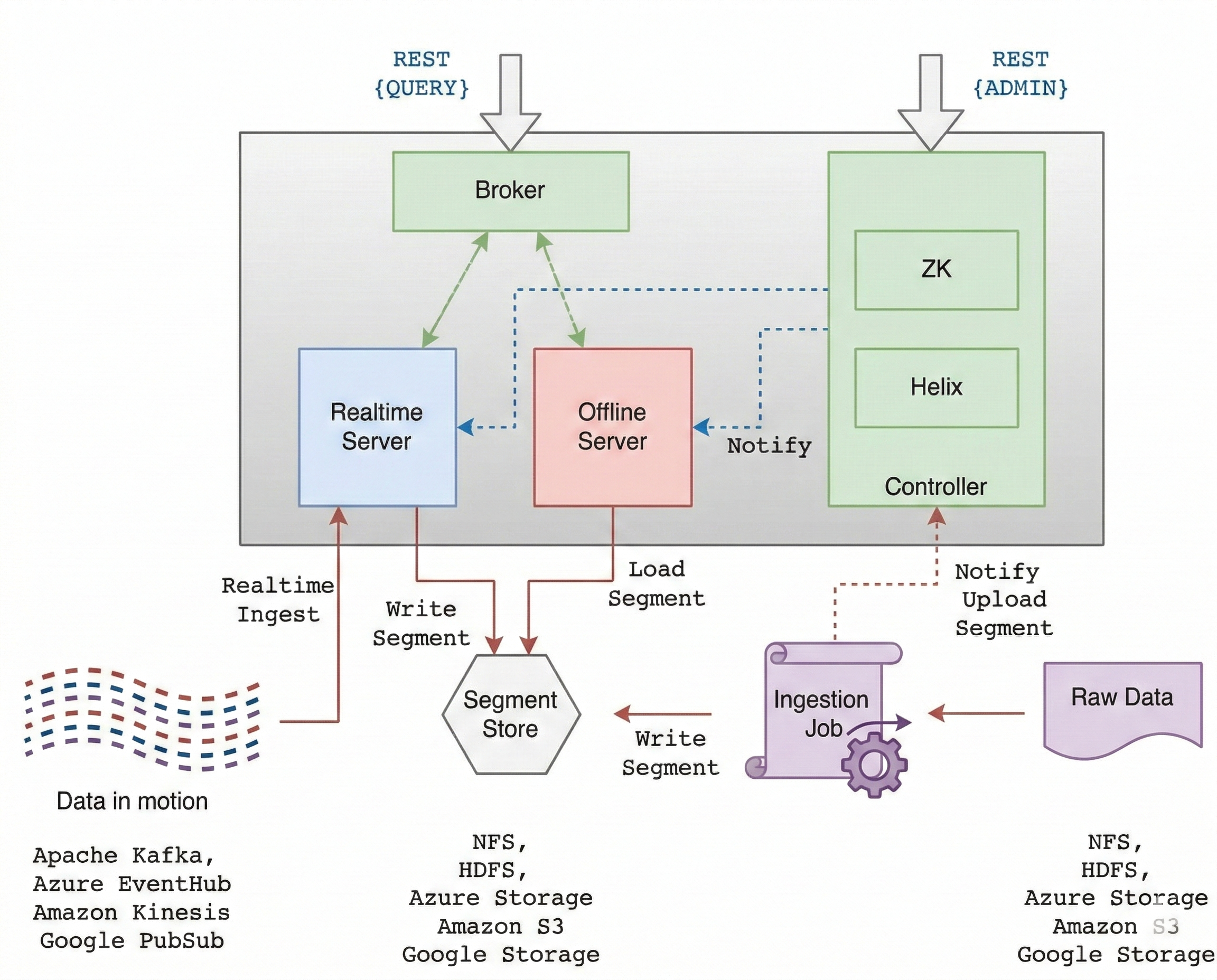}
  \caption{(Simplified) Pinot architecture.}
  \Description{Block diagram with controller, brokers, servers, segment store, and minions; arrows show query routing from broker to servers and metadata coordination from controller.}
  \label{fig:pinot-arch}
\end{figure}

\begin{enumerate}
  \item \textit{Controller.} Manages administrative tasks such as adding, deleting, and configuring Pinot tables and their segments. The controller maintains segment-to-instance mappings in the metadata store and coordinates cluster-wide operations.
  
  \item \textit{Broker.} The entry point for Pinot queries. Brokers perform scatter–gather execution by selecting target servers using routing tables, dispatching sub-queries in parallel, and aggregating/merging partial results into a final response returned to the client.
  
  \item \textit{Server.} Hosts segments (Pinot's term for shards) and executes queries. Servers perform the heavy lifting in query processing, scanning indexed data, applying filters, and computing partial aggregations that are sent back to brokers.
\end{enumerate}

Pinot has a flexible, horizontally scalable architecture that allows service providers to tune hardware for particular applications according to constraints of cost, maximum tolerated latency at a desired throughput, data retention, and schema.

\subsection{Background}
\label{sec:adss-background}
This section provides a technical background on Pinot that is necessary to understand the new design in the following sections. We will refer back to these concepts frequently.

\begin{figure}
  \centering
  \includegraphics[width=0.8\linewidth]{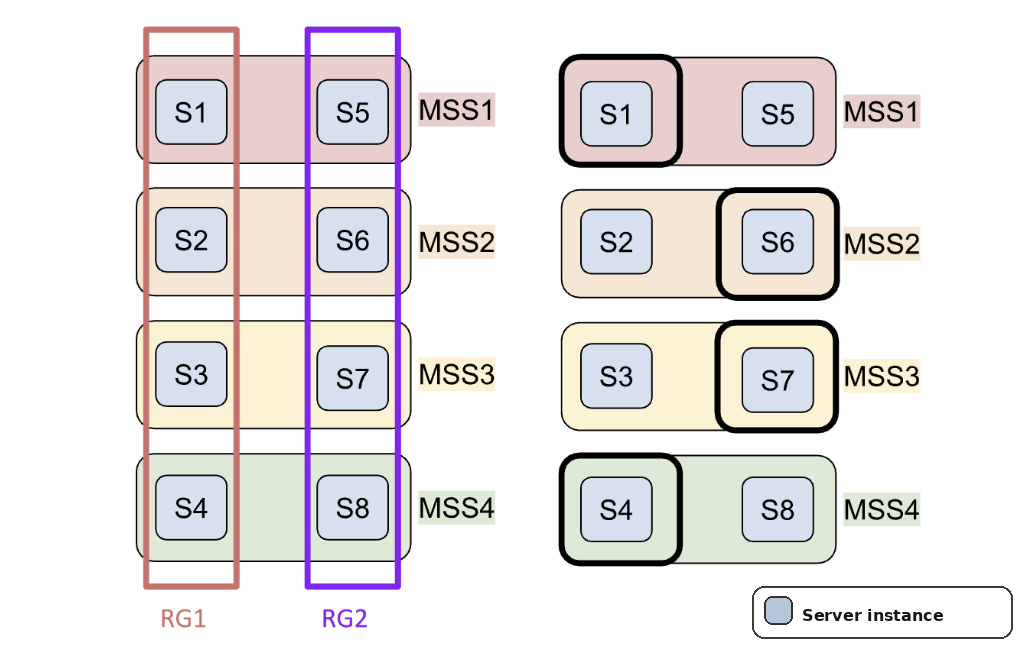}
  \caption{Replica group assignment and routing}
  \label{fig:routing}
\end{figure}

\textbf{Segment Assignment Strategies.} Most data systems use consistent hashing for partition assignment; CRUSH~\cite{weil2006crush} and CRUSHED~\cite{ApacheHelixCRUSHED} add topology-aware constraints but do not ensure fully even distribution or support capacity-based constraints. Helix~\cite{helix_waged_rebalancer} combines hard placement rules with soft resource constraints, but lacks assignment uniformity and rebalancing efficiency at Pinot’s scale.

Pinot uses replica-group segment assignment. The controller maintains a static segment-to-server mapping: it assigns all segments to servers in a designated replica group, then mirrors this across the remaining groups, ensuring every segment is replicated once per group on different servers (Figure~\ref{fig:routing}, left). The layout is symmetric across replica groups (RGs). We define a \emph{Mirror Server Set} (MSS) as the set of servers---one from each RG---hosting replicas of the same segment set. This mirrored structure enables seamless scaling, improved observability, A/B testing, and simpler operations.

\textbf{Routing Strategies.} With $N$ replica groups of $M$ servers each, the broker picks a replica group at random for each query and fans out sub-queries to the $M$ servers within it. Each group is a self-contained replica containing all segments, and the mirrored assignment ensures each server has an identical mirror in every other group. This allows Pinot to scale throughput (more replicas) and data size (more servers per replica) independently. If a server is unavailable (e.g., during deployment), the broker substitutes the corresponding mirror server from another group, keeping load reasonably balanced.

\subsection{Deployment Scale at LinkedIn}
\label{sec:deployment-scale}

All scale figures in this paper are for an entire fleet within a data center; LinkedIn operates multiple such data centers. Table~\ref{tab:deployment-scale} summarizes the fleet. The mechanisms described in this paper are deployed at varying scopes within this fleet in each data center.

\begin{table}[h]
  \centering
  \small
  \setlength{\tabcolsep}{6pt}
  \begin{tabular}{lr}
    \toprule
    \textbf{Metric} & \textbf{Value} \\
    \midrule
    Clusters                             & 250$+$                      \\
    Hosts (servers + brokers)            & 5{,}500$+$                  \\
    Tables                               & 4{,}500$+$                  \\
    Total data managed                   & $\sim$10\,PB (uncompressed) \\
    Fleet-wide query throughput          & 250{,}000$+$ QPS            \\
    P99 latency target (site-facing)     & 10--500\,ms                 \\
    P99 latency target (non-site-facing) & 500\,ms--1\,s               \\
    \bottomrule
  \end{tabular}
  \caption{LinkedIn's Pinot deployment at a glance (per datacenter fleet)}
  \label{tab:deployment-scale}
\end{table}

\subsection{Contributions}
\label{sec:contributions}
The primary contribution of this work is the design, implementation, and operationalization of a layered resiliency framework for large-scale real-time OLAP deployments. This framework addresses three critical vectors (workload, structural / topology, and runtime) of resiliency in distributed systems within a single architectural blueprint built into the internals of the core OLAP engine. Specifically, our contributions include:

\textbf{Query Workload Isolation (QWI).} OLAP engines on shared
  clusters must proactively manage queries that risk excessive memory
  or CPU use. Prior
  work~\cite{druid_cluster_tuning,apache-doris-wlg,starrocks-resource-group}
  either does not adaptively terminate queries in-flight or relies on
  process-level cgroup isolation~\cite{linux-cgroups-v2}, neither of
  which fits Pinot's shared-JVM execution model. We close this gap
  with an in-JVM framework for sampling-based per-thread accounting
  and end-to-end workload budgeting across brokers and servers,
  validated across 10 multi-tenant clusters and 1{,}500+ hosts,
  delivering fine-grained per-query attribution at sub-1\% overhead.

\textbf{Zone-Aware Placement and Rebalancing.} Pinot's mirrored
  replica groups impose a stricter constraint than CRUSHED-style
  topology-aware assignment~\cite{LiCRUSHEDBlog,ApacheHelixCRUSHED}:
  placement repair must preserve Mirror Server Sets, accommodate
  replacement nodes from arbitrary maintenance zones, and minimize data
  movement. We present a greedy swap algorithm that incrementally
  repairs MZ balance under continuous churn with provable convergence,
  paired with an impact-free rebalance procedure that drains query
  traffic before substantial data movement. Together, they completed a
  zero-downtime migration of 10\,PB across 5{,}500+ hosts and have
  sustained 99.9\% availability under up to 7\% daily node churn.

\textbf{Adaptive Server Selection (ADSS).}
Query routing should dynamically choose hosts based on cluster conditions rather than relying on round-robin assignments. We adapted prior adaptive-routing methodology~\cite{suresh2015c3,spotify_els_part2,finagle_peak_ewma} to Pinot's mirrored replica-group placement scheme by scoring servers within each Mirror Server Set using broker-local in-flight and latency signals. We further improve routing stability with softmax-based probabilistic selection, and use exhaustive simulation to study stability and fix edge cases under diverse workload and failure conditions.


Together, these mechanisms---workload isolation, impact-free rebalancing, zone-aware placement, and adaptive routing---provide holistic resiliency for production OLAP deployments at cloud scale.

\section{Query Workload Isolation}
\label{sec:QWI}

Modern OLAP platforms like Pinot run hundreds of heterogeneous workloads---interactive dashboards, recommendation pipelines, and near-real-time metrics---on shared compute clusters. Each workload has distinct query shapes, concurrency profiles, and latency SLOs, yet all contend for the same CPU and memory resources. This section presents Query Workload Isolation (QWI), a system that introduces workload-level resource budgeting into Pinot's query execution layer.

\subsection{Motivation}

One-cluster-per-workload is impractical: traffic is bursty and diurnal, ingestion would be duplicated across silos, and operational effort scales with cluster count. Shared pools exploit statistical multiplexing to raise utilization; the challenge becomes \emph{safe sharing}.

Coarse controls fail because per-query cost is heavy-tailed: a microsecond aggregation and a multi-second high-cardinality join are both ``one query'' to QPS caps and concurrency limits. Even with conservative sizing, we observe a persistent mismatch between cluster and workload health: aggregate CPU remains modest while workloads miss SLOs. The culprit is interference---short-lived surges from expensive workloads monopolize resources on hosts they land on, starving co-tenants. Secondary effects (head-of-line blocking, retry storms, GC churn) amplify tail latency.

These observations motivate QWI, which addresses resource management at two levels: (1) \emph{query-level accounting and killing} to handle runaway individual queries, and (2) \emph{workload-level budgeting} to enforce fair sharing across logical workload groups.

\subsection{Design Overview}

QWI models each workload as a first-class entity with enforceable CPU and memory budgets propagated across brokers and servers. The design satisfies four constraints:
\begin{enumerate}
  \item End-to-end enforcement: isolation spans both tiers with consistent semantics.
  \item Low overhead: less than 1\% CPU and memory even at high QPS.
  \item Local decisions: sub-millisecond responsiveness.
  \item Robustness under churn: tolerates failures without violating budget semantics.
\end{enumerate}

The architecture has four components. \emph{Query Workload Config} stores per-workload budgets in ZooKeeper. \emph{Budget Manager} maintains a per-host ledger tracking remaining CPU and memory for each workload. The \emph{Resource Accountant} continuously measures per-query consumption using lock-free sampling. The \emph{Budget Enforcer} applies admission control and in-execution enforcement.

\subsection{Sampling-based Resource Accounting}\label{sec:accounting}

The foundation of QWI is accurate, low-overhead measurement of per-query resource consumption. This is challenging in Pinot's execution model: queries decompose into tasks distributed across thread pools, with each thread processing multiple segments. Combined with \textit{Java}'s opaque memory management, attributing usage to individual queries is difficult.

Existing approaches are not directly applicable. Trino's allocation-time accounting blocks when memory is insufficient, conflicting with Pinot's millisecond SLAs. Heuristic approaches (Spark, Lucene) require per-data-structure tuning, creating a maintenance burden and extra effort to retrofit each existing customize data structure.

\begin{algorithm}[!h]
\footnotesize
\caption{Sample And Aggregate Metrics}
\label{alg:sampleAndAggregate}
\begin{algorithmic}[1]
\State {\em // T: execution threads;\ \ S(tid): current $\langle qId, tId\rangle$;\ \ M(tid): metric}
\State {\em // previousTask, activeMetric: per-thread state;\ \ inactiveMetric: per-query}
\Procedure{sampleAndAggregate}{T}
\State activeQueries $\gets \emptyset$
\For{t $\in$ T} \Comment{sample each running thread}
    \State $\langle qId, tId\rangle \gets$ S(t.id);\ \ metric $\gets$ M(t.id)
    \State activeQueries $\gets$ activeQueries $\cup \{qId\}$
    \State $\langle qId', tId'\rangle \gets$ previousTask(t.id)
    \If{$\langle qId', tId'\rangle = \langle qId, tId\rangle$}
        \State activeMetric(t.id) $\gets$ metric
    \Else \Comment{task switched: flush prior task}
        \State inactiveMetric($qId'$) $\mathrel{+}=$ activeMetric(t.id)
        \State previousTask(t.id) $\gets \langle qId, tId\rangle$
        \State activeMetric(t.id) $\gets$ metric
    \EndIf
\EndFor
\If{usageThresholdCheck()} \Comment{aggregate per-query for action}
    \State activeQueryToUsage $\gets$ \{\}
    \For{$qId \in$ activeQueries}
        \State activeQueryToUsage($qId$) $\mathrel{+}=$ inactiveMetric($qId$)
    \EndFor
    \For{t $\in$ T}
        \State $\langle qId, \_\rangle \gets$ previousTask(t.id)
        \State activeQueryToUsage($qId$) $\mathrel{+}=$ activeMetric(t.id)
    \EndFor
    \State {\em // pluggable strategy uses activeQueryToUsage}
\EndIf
\For{$qId \in$ inactiveMetric \textbf{where} $qId \notin$ activeQueries}
    \State inactiveMetric.remove($qId$)
\EndFor
\EndProcedure
\end{algorithmic}
\end{algorithm}

\begin{figure}[h]
  \centering
  \includegraphics[width=1\linewidth]{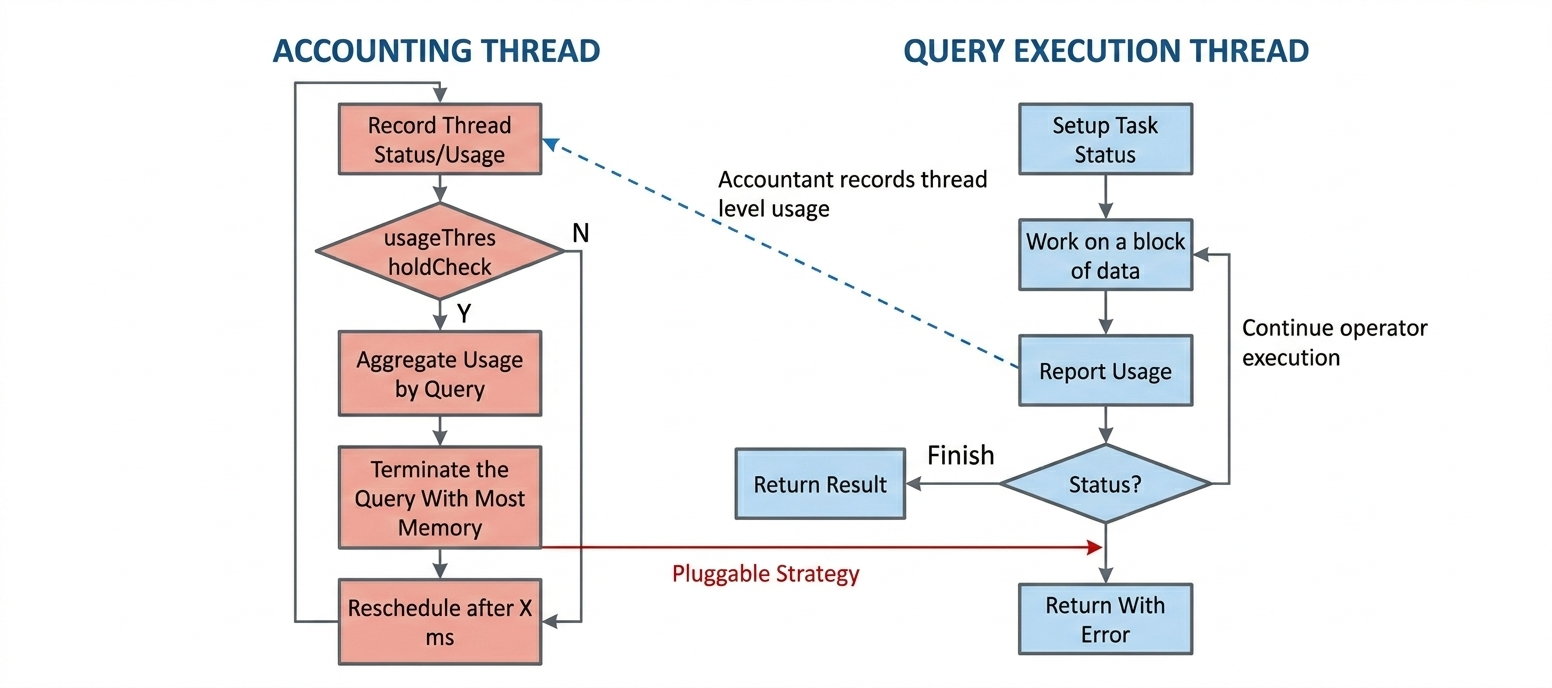}
  \caption{Resource accounting}
  \label{fig:accounting}
\end{figure}

We introduce a sampling-based accounting framework that separates metric reporting (per-thread, lock-free) from aggregation (centralized, periodic), as shown in Figure~\ref{fig:accounting}. Upon accepting a task, each worker thread establishes a context with a $\langle\text{queryID}, \text{taskID}\rangle$ pair. During task execution, before processing each sub-chunk of data (usually a few thousand records), the worker reports CPU time and memory metrics to a per-thread region using volatile primitives. Algorithm~\ref{alg:sampleAndAggregate} summarizes the sampler's aggregation logic. Periodically, a dedicated sampler scans all threads, tracking task transitions to determine active queries and their resource consumption. This design achieves less than 0.3\% CPU overhead at 500 QPS with a 1\,ms sampling interval and sub-0.1\,ms thread reporting interval, demonstrating fine-grained tracking without impacting latency. Specifically for memory, QWI samples the JVM's per-thread \texttt{getThreadAllocatedBytes} counter, which is cumulative rather than live retained heap; therefore, high-churn short-lived objects are still reflected in allocation accounting. This is intentional: allocation volume captures the GC pressure a query induces, not only its retained heap footprint, and both allocating and reclaiming memory consume shared system resources. Heap-pressure detection is handled separately using JVM-reported total heap usage, as described in Section~\ref{sec:enforcement}.


\subsection{Workload Configuration and Budget Propagation}

Operators define workloads with explicit, continuously enforced resource budgets so that no workload exceeds its fair share. Each workload has two per-window budgets: \texttt{cpuCostNs} (CPU time in ns, measured via \texttt{ThreadMXBean}) and \texttt{memoryCostBytes} (heap allocation, tracked via sampling). CPU time provides consistent per-thread measurement semantics within an enforcement domain, but it is not a hardware-normalized measure of work across CPU generations. In our deployment, QWI enforces budgets within homogeneous clusters; across clusters with different CPU models, operators calibrate workload budgets separately from observed workload behavior, allowing \texttt{cpuCostNs} values to scale with differences in IPC, scheduling, cache behavior, and pipeline efficiency. Automatic normalization across heterogeneous hardware pools is future work.

\begin{figure}[h]
\begin{lstlisting}[language=json, basicstyle=\ttfamily\tiny]
{
  "workloadName": "analytics-workload",
  "nodeConfigs": [
    {
      "nodeType": "SERVER",
      "enforcementProfile": {
        "cpuCostNs": 1.0e9,
        "memoryCostBytes": 5.0e9
      },
      "propagationScheme": {
        "type": "TABLE",
        "tables": ["tableA", "tableB"]
      }
    }
    ...
  ]
}
\end{lstlisting}
\caption{Example workload configuration with per-node resource budgets and table-level propagation.}
\label{fig:workload-config}
\end{figure}

Configurations specify budgets separately for brokers and servers (Figure~\ref{fig:workload-config}); the Pinot controller translates them into per-host limits. Propagation operates in two modes: \emph{table-level} pushes budgets to hosts serving specified tables, while \emph{tenant-level} propagates to all hosts in a tenant.



On each host, a \texttt{BudgetManager} tracks remaining budget per (workload, resource) and exposes a lock-free \texttt{tryCharge}. Budgets decrement monotonically within a window and reset on rollover.

\subsection{Enforcement}
\label{sec:enforcement}

QWI enforces resource limits at two complementary levels.

\textbf{Query-level enforcement.}
Individual queries that threaten system stability are terminated based on resource thresholds. OOM-risk detection uses JVM-reported total heap usage rather than the sum of sampled per-query allocations. If heap usage crosses configured thresholds (e.g., 96\%), QWI kills the most resource-intensive query; at 99\%, all queries are terminated to prevent OOM crashes. For latency-sensitive deployments, per-query CPU-time limits trigger earlier intervention. This provides a critical safety net: even without workload budgets configured, the system protects itself from runaway queries.

\textbf{Workload-level enforcement.}
Workload budgets enable proactive, continuous isolation. Each query is tagged with its workload, and resource consumption is charged against that workload's budget. 

The current design enforces strict per-workload budgets without support for bursting (temporarily exceeding budgets when cluster capacity is available) or quota stealing (borrowing unused budget from other workloads). These features are planned for future work, as they could improve utilization while requiring careful design to maintain isolation guarantees.

\begin{figure}[h]
    \centering
    \includegraphics[width=0.90\linewidth]{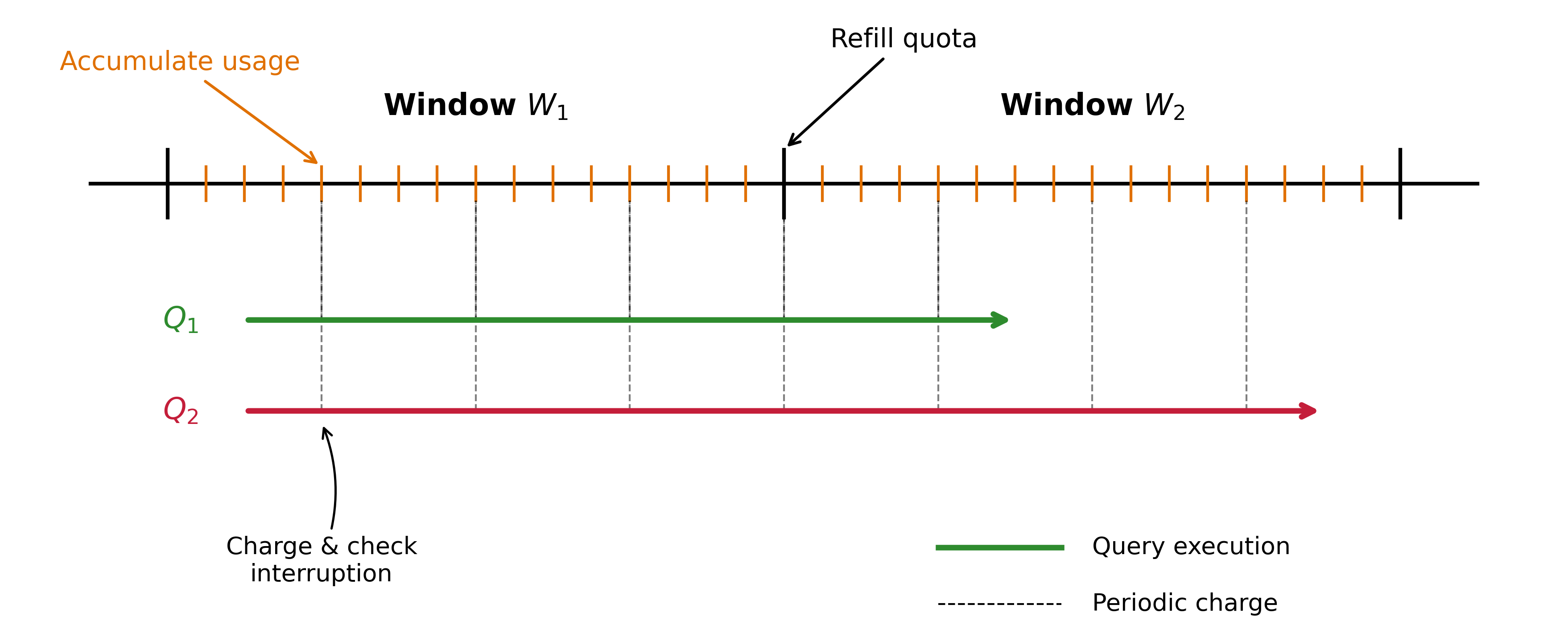}
    \caption{Enforcement timeline: budgets refill each window; queries charge against their workload's budget continuously.}
    \label{fig:enforcement}
\end{figure}

\begin{algorithm}[h]
\small
\caption{Per-Thread Budget Enforcement}
\label{alg:budgetEnforcement}
\begin{algorithmic}[1]
\State // {\em w: workload associated with query q}
\State // {\em CPU, MEM: resource types}
\Procedure{enforcebudget}{q, w}
\State prevCpu = threadCpuNow()
\State prevMem = threadBytesNow()
\For{each accounting interval}
    \State cpu = threadCpuNow()
    \State mem = threadBytesNow()
    \State cpuDelta = cpu - prevCpu
    \State memDelta = mem - prevMem
    \State {\em // Charge CPU first, then memory, If either fails, cancel query}
    \If{!tryCharge(w, CPU, cpuDelta) \textbf{or} !tryCharge(w, MEM, memDelta)}
        \State cancelQuery(q)
        \State \textbf{break}
    \EndIf
    \State prevCpu = cpu
    \State prevMem = mem
\EndFor
\EndProcedure
\end{algorithmic}
\end{algorithm}

These two mechanisms are complementary: query-level enforcement handles individual outliers and emergency scenarios, while workload-level enforcement ensures fair sharing under normal operation. Both operate host-locally, bounding blast radius and avoiding coordination overhead.

Workload-level enforcement operates at two stages of query processing. At admission, QWI performs a provisional charge against the workload's budget before query planning; if insufficient budget remains, the query is rejected immediately, preventing exhausted workloads from queuing additional work. During execution, worker threads periodically charge their measured CPU and memory deltas (Figure~\ref{fig:enforcement}); When a workload's budget reaches zero on a host, QWI (1) stops admitting new queries for that workload and (2) may preemptively cancel in-flight queries, depending on configuration. All enforcement is strictly host-local---each broker and server maintains its own view of workload budgets and makes decisions independently. This bounds blast radius (a budget-exhausted workload on one host doesn't affect other hosts) and avoids coordination overhead in the critical path.

The enforcement window length $W$ controls the tradeoff between responsiveness and stability: large windows behave like hard quotas that throttle until reset, while small windows act as rate limiters that smooth bursts. Our evaluation found that 5--10 second windows work best in production---shorter windows over-represent transient spikes when estimating p95-based budgets.

QWI is designed for independent operation under partial failures. Hosts cache their assigned budgets locally, so enforcement continues uninterrupted even if the controller becomes unavailable. When a host restarts or misses a propagation message, it fetches current budgets from the controller on startup; the idempotent \texttt{addOrUpdateWorkload} API ensures safe reapplication without risk of duplicate or stale state.

\subsection{Evaluation}

We evaluated QWI across five dimensions: runtime overhead, budget enforcement accuracy, workload isolation effectiveness and parameter sensitivity. Experiments ran on three production-grade clusters, each with approximately 50 instances on Linux servers with 40 vCPUs (Intel Xeon E5-2680 v3), 128\,GB RAM, and 3\,TB SSD:

\begin{enumerate}
  \item Cluster~A: Lightweight aggregations, high throughput (~1000 QPS/replica)
  \item Cluster~B: Compute-heavy group-by queries, moderate throughput (~100 QPS/replica)
  \item Cluster~C: Mixed query complexity, latency-sensitive (~30 QPS/replica)
\end{enumerate}

\begin{table}[h]
  \centering
  \small
  \setlength{\tabcolsep}{4pt}
  \begin{tabular}{lrrrrrr}
    \toprule
    & \multicolumn{2}{c}{CPU (\%)} &
      \multicolumn{2}{c}{Heap (GB)} &
      \multicolumn{2}{c}{P99 (ms)} \\
    \cmidrule(lr){2-3} \cmidrule(lr){4-5} \cmidrule(lr){6-7}
    Cluster & Base & +QWI & Base & +QWI & Base & +QWI \\
    \midrule
    A & 12.3 & 12.4 & 18.0 & 18.4 & 45 & 45 \\
    B & 21.8 & 22.0 & 18.5 & 19.0 & 65 & 66 \\
    C & 30.0 & 30.3 & 19.8 & 20.3 & 38 & 38 \\
    \bottomrule
  \end{tabular}
  \caption{Runtime overhead: $<$1\% CPU, 0.5\,GB heap, no P99 impact.}
  \label{tab:overhead}
\end{table}

\textbf{Overhead.} A practical isolation mechanism must impose minimal overhead on normal query execution. We use the three clusters above for controlled trace replay because they allow direct baseline comparison against the same workload with and without QWI enabled; deployment-scale evidence is reported separately below. We replayed 24-hour production traces against each cluster. As shown in Table~\ref{tab:overhead}, QWI added less than 1\% CPU overhead across all cluster types
from high-throughput lightweight queries (Cluster~A) to compute-intensive workloads (Cluster~B). Heap usage increased by approximately 0.5\,GB due to per-query tracking structures. Critically, P99 latency remained unchanged, confirming that QWI's sampling-based accounting does not introduce latency variability into the query path.

\textbf{Production scale.} Following the controlled evaluation above, we have rolled out QWI's resource accounting and cost collection to 10 of the 250+ production clusters (Table~\ref{tab:deployment-scale}) comprising 1{,}500+ hosts (servers and brokers). These clusters serve as a litmus test for QWI: by hosting workloads with widely varying query complexity, concurrency, and latency SLOs on shared hardware, they expose the noisy-neighbor interference that QWI is designed to prevent. Per-host overhead and accounting accuracy remain consistent with the controlled results. Cost collection and enforcement share the same hot path, so the measured overhead is representative; budget enforcement based on these signals is currently being rolled out incrementally. This matches QWI's scalability model: admission, accounting, and enforcement are host-local, so each broker and server makes decisions against its own cached budgets, and the query-path cost does not grow with cluster size. Controller-side cost for workload configuration changes is also below 1\% of controller capacity, since these events occur at adhoc-driven rates (under 1QPS in our deployment) rather than on the query path.

\begin{figure}[h]
  \centering
  \begin{subfigure}[b]{0.49\linewidth}
    \includegraphics[width=\linewidth]{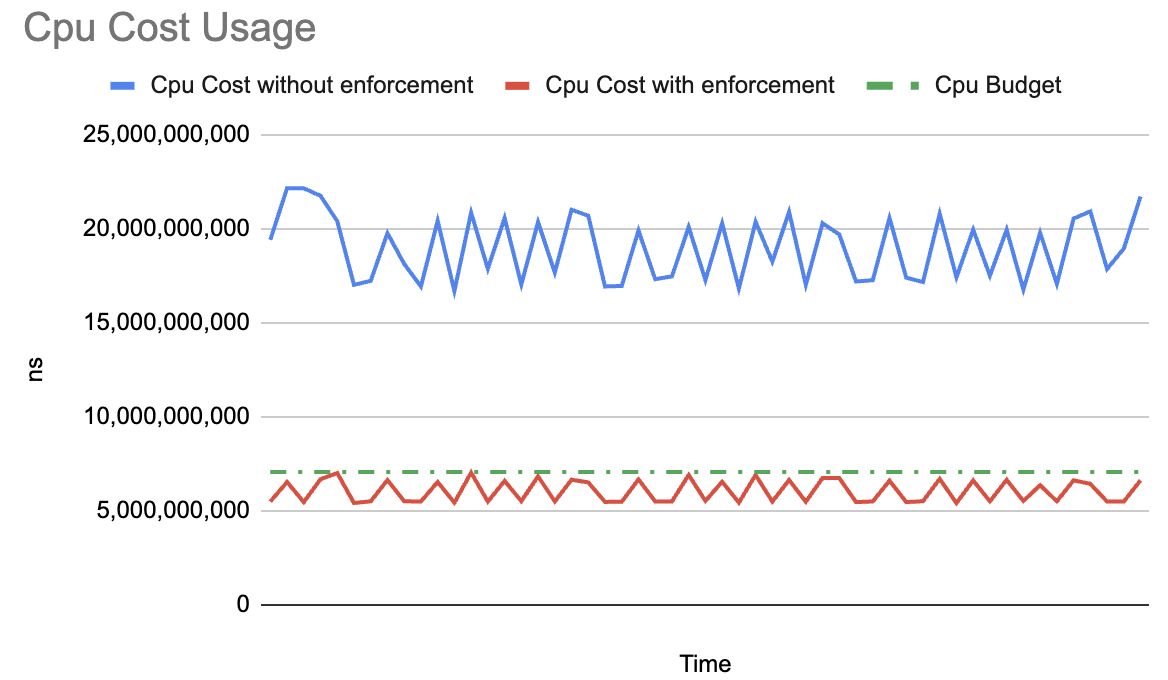}
    \caption{CPU cost}
  \end{subfigure}
  \begin{subfigure}[b]{0.49\linewidth}
    \includegraphics[width=\linewidth]{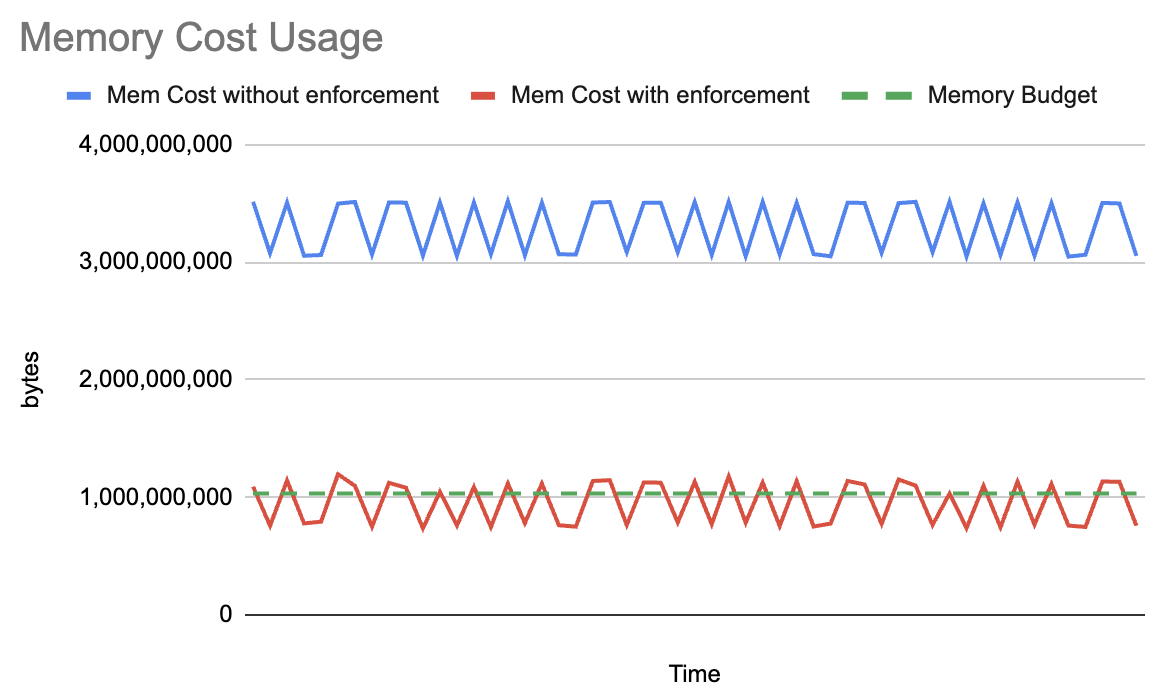}
    \caption{Memory cost}
  \end{subfigure}
  \caption{Budget enforcement: without QWI (blue), cost scales with QPS; with QWI (red), cost flattens at budget (green).}
  \label{fig:enforcement-result}
\end{figure}

\textbf{Budget enforcement.} We evaluated whether QWI accurately enforces configured budgets under load increases. For each workload, we set CPU and memory budgets to match baseline usage, then gradually increased query volume by 10--50\%. Figure~\ref{fig:enforcement-result} shows representative results from Cluster~A. Without QWI (blue line), resource consumption scaled proportionally with QPS---a 50\% traffic increase produced approximately 50\% higher CPU and memory usage. With QWI enabled (red line), consumption flattened at the configured budget (green line) as excess queries were rejected. Across all clusters, CPU budget accuracy ranged from 95--100\%, while memory accuracy ranged from 90--100\%. The slightly lower memory accuracy reflects the inherent variability in per-query allocation patterns and the sampling-based measurement approach.

\begin{figure}[h]
  \centering
  \includegraphics[width=0.9\linewidth]{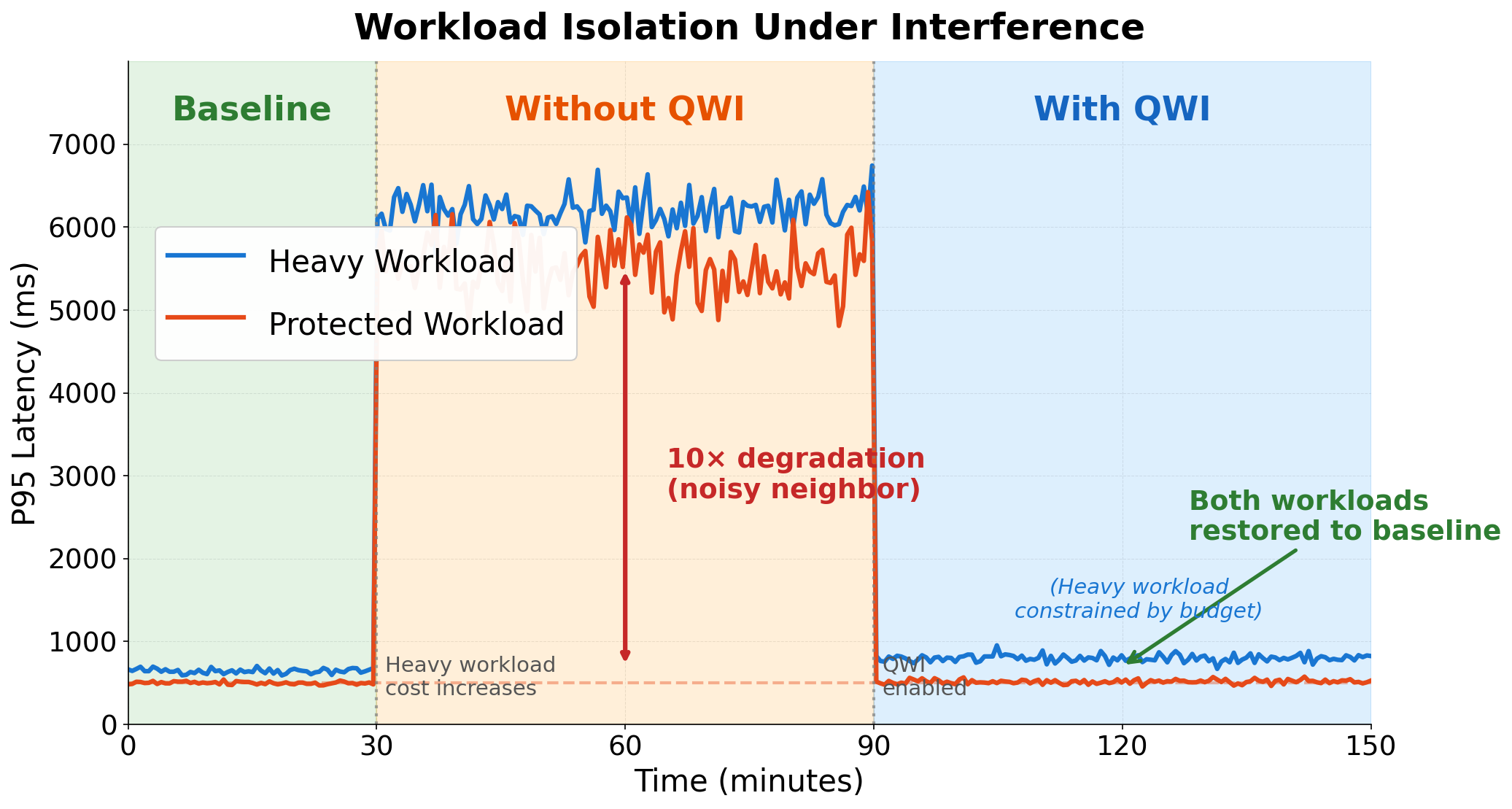}
  \caption{Workload isolation: heavy workload degrades co-tenant P95 without QWI; with QWI, co-tenant recovers.}
  \label{fig:isolation}
\end{figure}

\textbf{Workload isolation.} The critical test for QWI is whether it protects co-located workloads from interference---the noisy-neighbor problem that plagues multi-tenant OLAP deployments. We ran two workloads on shared infrastructure: a compute-heavy workload (Workload~1) and a latency-sensitive workload (Workload~2). Figure~\ref{fig:isolation} shows P95 latency across three phases:

\begin{enumerate}
  \item \emph{Baseline}: Both workloads operate at steady state with P95 latencies of approximately 650\,ms and 500\,ms respectively, well within their SLA targets.
  \item \emph{Without QWI}: We increased Workload~1's query complexity by introducing more expensive aggregation patterns. Both workloads' P95 latencies spiked to over 6 seconds---a 10$\times$ degradation---even though Workload~2's traffic pattern was unchanged. This demonstrates the noisy-neighbor problem: resource contention from one misbehaving workload propagates to starve co-tenants of CPU and memory.
  \item \emph{With QWI}: Under identical conditions, Workload~2's P95 returned to baseline (approximately 500\,ms with less than 5\% deviation) while Workload~1 was constrained to its configured budget. Excess queries from Workload~1 were rejected rather than allowed to consume shared resources.
\end{enumerate}

These results validate QWI's core value proposition: workloads can safely share infrastructure without risking SLA violations from neighboring workloads. The protected workload experiences consistent latency regardless of co-tenant behavior.

\begin{figure}[h]
  \centering
  \begin{minipage}[b]{0.49\linewidth}
    \centering
    \includegraphics[width=\linewidth]{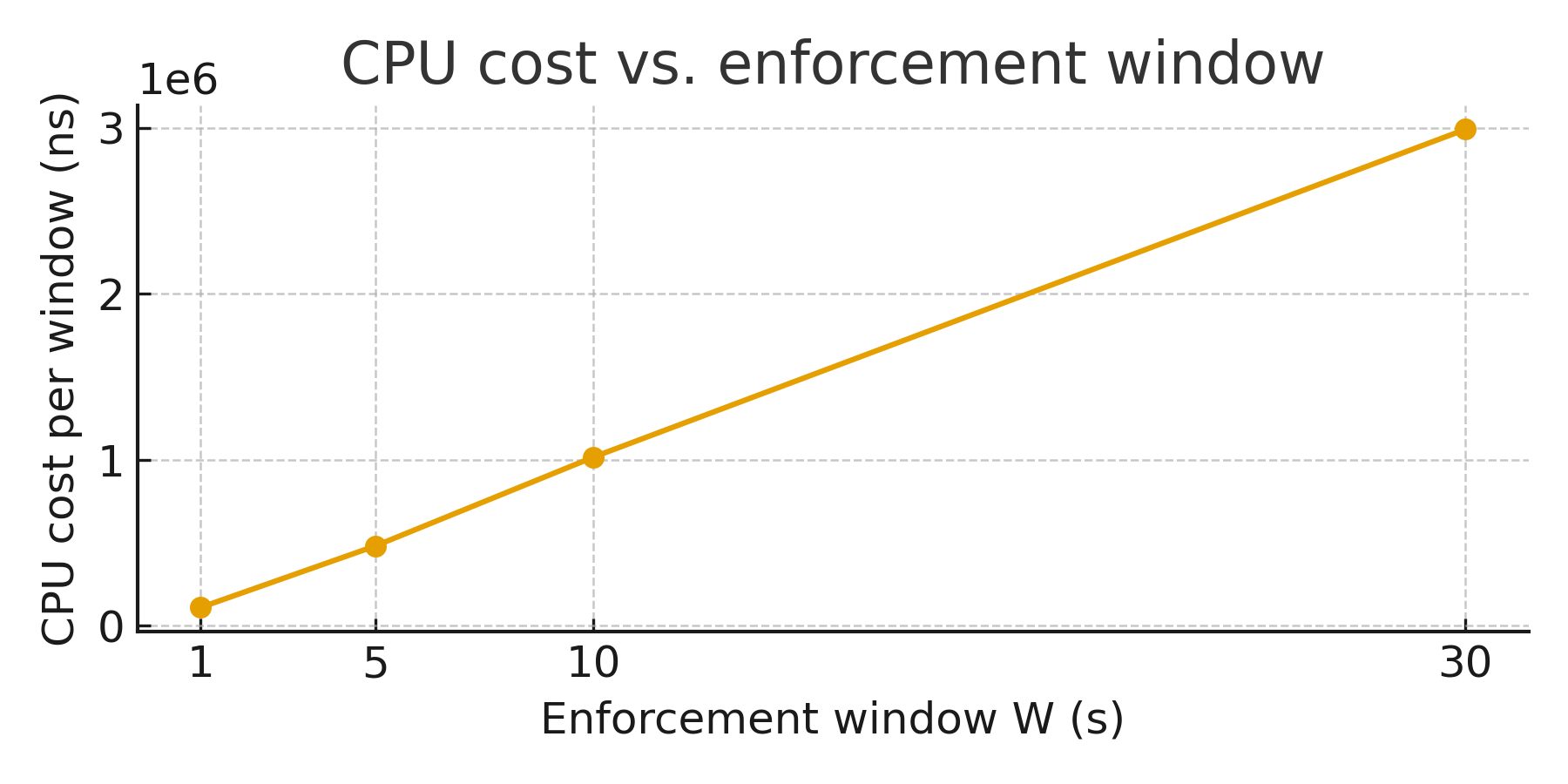}
    \centerline{(a) CPU vs. window}
  \end{minipage}
  \hfill
  \begin{minipage}[b]{0.49\linewidth}
    \centering
    \includegraphics[width=\linewidth]{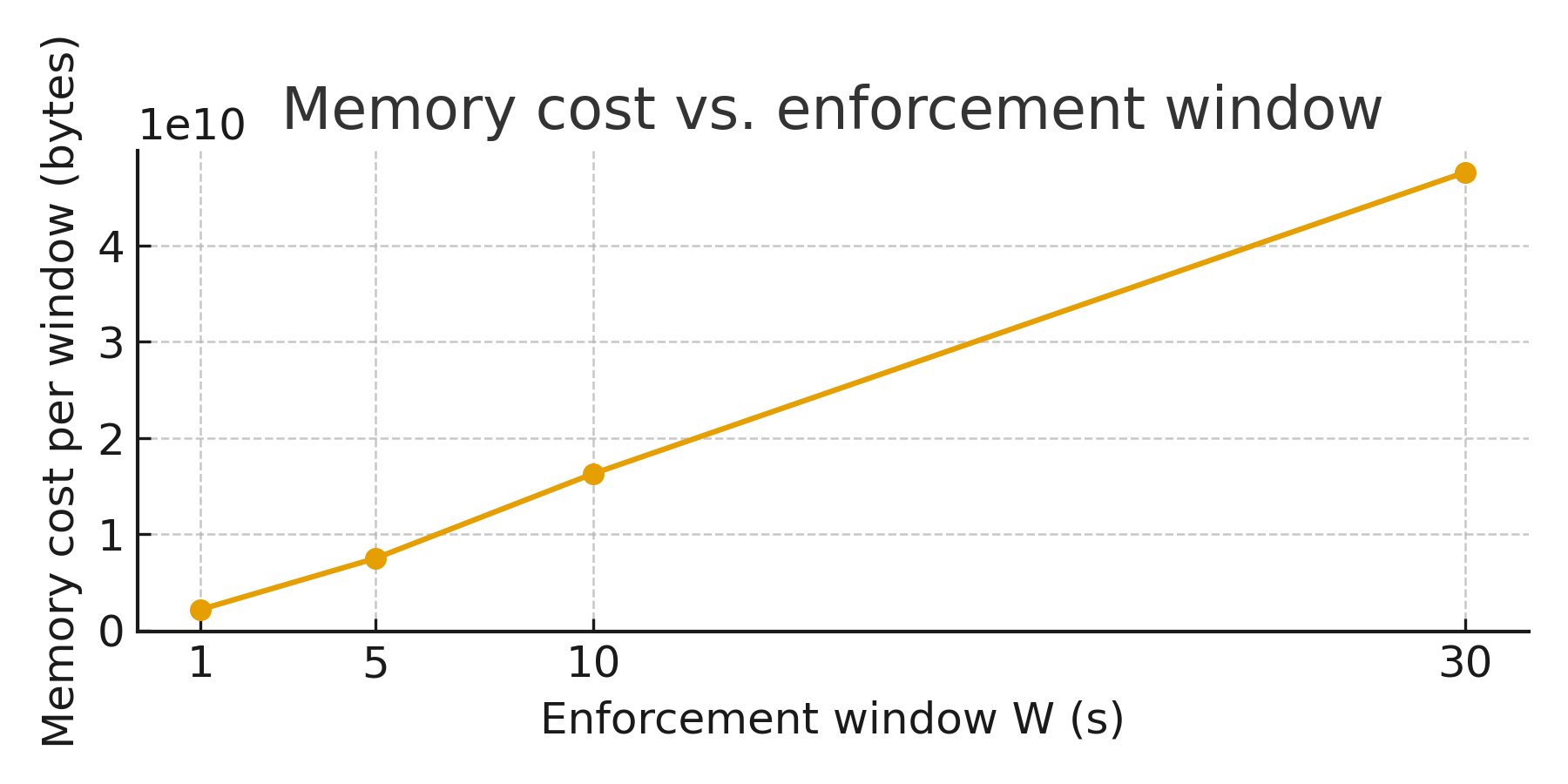}
    \centerline{(b) Memory vs. window}
  \end{minipage}
  \caption{Window sensitivity: cost scales linearly with window length.}
  \label{fig:window}
\end{figure}

\textbf{Window sensitivity.}
The enforcement window $W$ governs the responsiveness-stability tradeoff. Figure~\ref{fig:window} shows that CPU and memory costs scale linearly with window length, confirming unbiased accounting. We recommend 5-second windows for production: shorter windows over-represent transient spikes in p95-based budgets, while longer windows delay response to sustained overload.

\begin{figure}[h]
  \centering
  \includegraphics[width=0.85\linewidth]{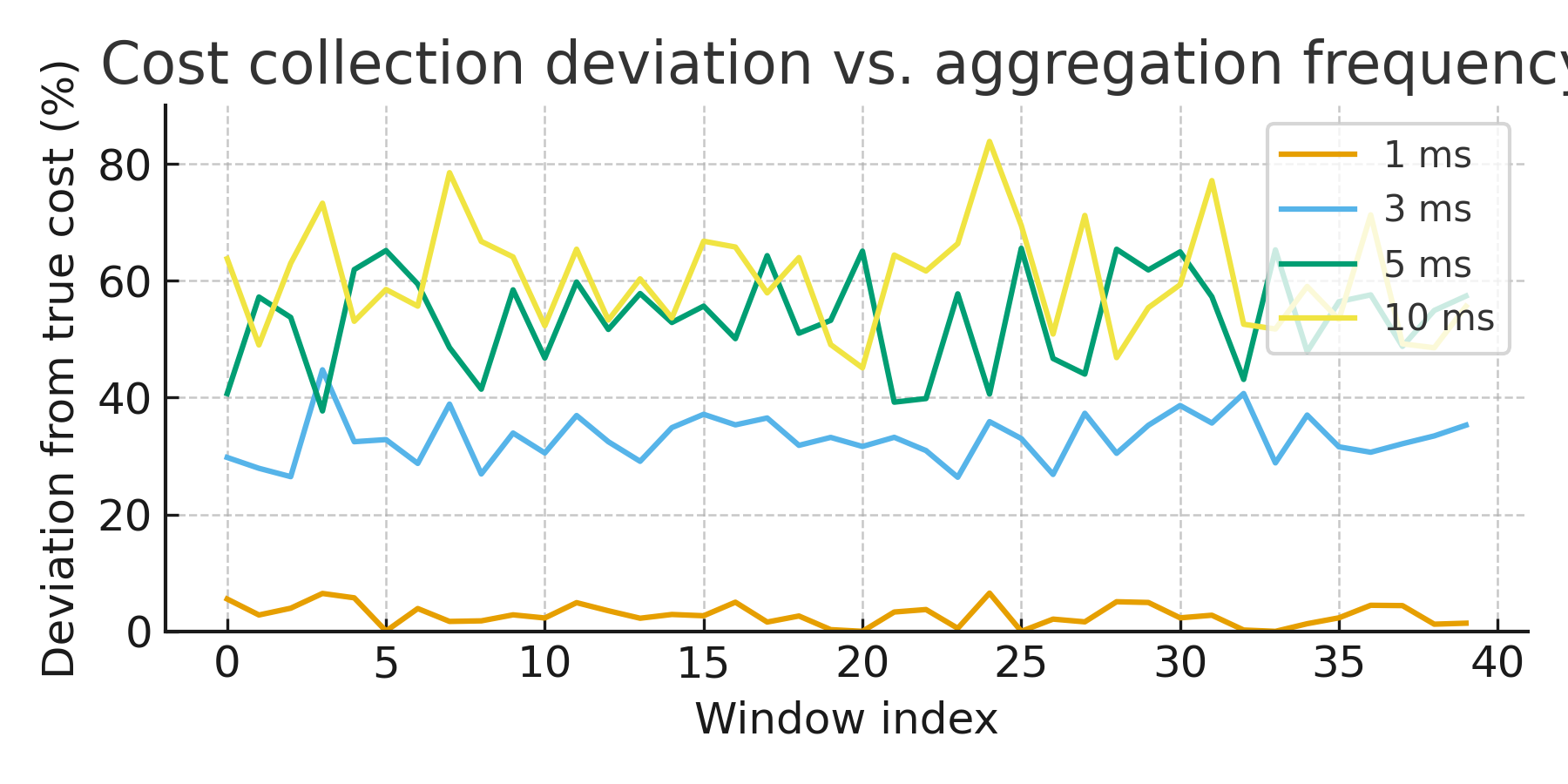}
  \caption{Aggregation frequency: 1\,ms achieves near-zero deviation from ground truth.}
  \label{fig:aggfreq}
\end{figure}

\textbf{Aggregation frequency.} Sampling frequency significantly affects measurement accuracy (Figure~\ref{fig:aggfreq}). At 1\,ms intervals, deviation from ground truth is near zero; at 5--10\,ms, errors reach 50--80\% as short-lived queries complete between samples. The 1\,ms default captures fine-grained behavior while adding less than 1\% overhead.

\subsection{Lessons Learned}
\textbf{Interaction with GC.} Heap-pressure enforcement interacts with JVM garbage collection policy because QWI observes JVM-reported total heap usage. In production,  G1GC sometimes left reclaimable objects visible long enough for heap-pressure detection to kill queries unnecessarily. We therefore made a decision to switch to concurrent gc which more frequently sweep out the unreachable objects with shorter pauses. We first experimented with Shenandoah GC and set the critical query-killing threshold to 96\%, above Shenandoah's default GC trigger point (roughly 90\% heap occupancy), allowing GC to reclaim memory before QWI treats pressure as sustained. This has eliminated all false positives we saw. However, we ran into an edge case where Shanadoah GC's heap reporting is delayed at high load. The debugging effort did not yield a clear root cause. Therefore, we are on our path to switch to ZGC.

\noindent\textbf{Workload-aware, measurement-driven isolation.} Production query isolation must be both workload-aware and measurement-driven: table- and tenant-level isolation are too coarse, and analytical query cost is hard to predict before execution, which is why we charge measured resources during execution rather than estimated cost at admission. Safe adoption also required a staged rollout — collect cost first, validate budgets against historical traffic with headroom, then enable enforcement — so calibration errors surface as observability gaps rather than false positives. A direction for future work is explicit modeling of workload drift, since a workload can grow more expensive at constant QPS as query patterns, cardinality, or data layout shift.

\section{Zone-Aware Placement and Rebalancing}
\label{sec:rebalancing}

 \subsection{Motivation}

  Pinot clusters frequently undergo lifecycle events---scaling, rolling
  upgrades, kernel patches, and failure recovery---each requiring data
  redistribution (\emph{rebalancing}). This raises two interrelated
  challenges.

  First, the redistribution itself degrades query performance: data
  downloads consume network and disk bandwidth, disk writes evict the
  page cache, index generation on newly loaded segments is CPU-intensive,
  and segment load becomes unbalanced while data is in motion.

  Second, replicas must be placed with awareness of fault-domain
  topology. We refer to fault domains as \emph{Maintenance Zones} (MZs)
  at LinkedIn---logical groupings of nodes, racks, power circuits, or
  availability zones that share a common failure
  risk~\cite{aws-placement-groups,azure_az_overview,hadoop_rack_awareness}.
  Operators drain one MZ at a time during planned maintenance; cluster
  managers spread workloads across MZs using topology-aware
  constraints~\cite{k8s_topology_spread}. Prior Helix placement
  work~\cite{LiCRUSHEDBlog} supports topology-aware assignment but does
  not fit Pinot's replica-group-based assignment and routing
  (Section~\ref{sec:adss-background}) or migration constraints
  (Section~\ref{sec:relatedFd}).

  We address both challenges (Sections~\ref{sec:mz-aware}
  and~\ref{sec:impact-free-rebalance}): an MZ-aware assignment algorithm
  that maximizes replica diversity across fault domains, and an
  impact-free rebalance procedure that converges to the target
  assignment without degrading SLAs.

\subsection{Maintenance Zone Aware Assignment}
\label{sec:mz-aware}

We assume the following platform properties~\cite{azure_vmss_auto_repairs,azure_vmss_scale_in_policy}:

\begin{enumerate}
    \item \label{mzassume:1} Pods are provisioned from a fixed number of MZs; an even distribution across MZs is maintained.
    \item When a node becomes unhealthy, its pods are evacuated and rescheduled~\cite{k8s_pod_lifecycle,azure_vmss_auto_repairs}. When using local disks, node loss can drop local state (e.g., NVMe SSD), causing temporary unavailability during bootstrap~\cite{k8s_local_ephemeral_storage}. The new node's MZ is not guaranteed to match the old one, but the distribution in Assumption~\ref{mzassume:1} is assumed to hold.
    \item During cluster uplift, newly added nodes are spawned in (effectively) random MZs.
    \item During cluster downlift, the operator specifies the number of pods/VMs to remove rather than specific nodes; removals may be effectively random~\cite{azure_vmss_scale_in_policy}.
\end{enumerate}

Under these assumptions, a fixed mapping between replica groups and MZs is clearly too rigid. We therefore aim to design a data segment assignment algorithm that satisfies the following goals:

\begin{enumerate}
    \item \label{mzgoal:1} All assignments generated by the algorithm should achieve best-effort replica spreading across MZs, such that when one MZ is taken out for maintenance, the impact on data segment availability is minimized.

    \item Any assignment change during the cluster lifecycle---i.e., initial migration, node swap, uplift, or downlift---should minimally perturb the existing assignment. This reduces the cost of data movement and transmission, as well as any capacity loss incurred during rebalancing.

    \item The operational model must be easy to operate and debug.
\end{enumerate}

\textbf{Algorithm Design.}
As discussed in Section~\ref{sec:adss-background}, Pinot stores a static assignment map in the metadata store to perform instance assignment, as illustrated in Figure~\ref{fig:routing}. In this representation, Pinot explicitly maintains the mapping for one replica of each segment and keeps the other replicas mirrored. To recap, we refer to each column---a set of machines containing all segments---as a \emph{replica group}, and each row---a set of machines hosting exactly the same set of segments---as a \emph{mirrored server set}.

To translate the design goals into a formal specification, we impose the following constraints:

\begin{enumerate}
    \item \label{mzconstraint:1} When one MZ is down (due to maintenance, power outage, etc.), for each segment:
        \begin{enumerate}
            \item When $R \le MZ$, no more than one replica of that segment goes down.

            \item {When $R > MZ$} no more than $\lceil R / MZ \rceil$ replicas go down.
        \end{enumerate}
    Here, we denote the number of MZs by $MZ$ and the number of replica groups by $R$.

    \item After scaling or node swaps occur, the number of segments that must be moved to satisfy Constraint~\ref{mzconstraint:1} should be minimized.
\end{enumerate}

\begin{algorithm}[!h]
\small
\caption{Swap to Maintenance Zone Aware}
\label{alg:swaptomzaware}


\begin{algorithmic}[1]
\Procedure{SwapToMZAware}{$S$}
    \For{$s \in S$}
        \While{$\text{dup}(s) > 0$}
            \For{$x \in s$ \textbf{where} $\text{isDuplicateMZ}(x)$}
                \If{\Call{TrySwap}{$s, x, S$}}
                    \State \textbf{break}
                \EndIf
            \EndFor
        \EndWhile
    \EndFor
\EndProcedure

\Function{TrySwap}{$s, x, S$}
    \For{$s' \in S \setminus \{s\}$}
        \For{$x' \in s'$}
            \State $m, n \gets \text{dup}(s),\ \text{dup}(s')$
            \State $\text{swap}(x', x)$
            \If{$\text{dup}(s) + \text{dup}(s') < m + n$}
                \State \Return \textbf{true}
            \EndIf
            \State $\text{swap}(x, x')$
        \EndFor
    \EndFor
    \State \Return \textbf{false}
\EndFunction
\end{algorithmic}
\end{algorithm}

Algorithm~\ref{alg:swaptomzaware} presents the greedy algorithm. We assume the system starts in a \emph{good} state satisfying Constraint~\ref{mzconstraint:1}. With balanced nodes across MZs, this can be built by filling each mirrored server set round-robin over MZs. The challenge arises when node swaps or scaling introduce servers in random MZs: multiple instances of the same mirrored server set may land in one MZ (Eq.~\ref{eq:bad-row}), so draining that zone can exceed tolerated capacity.

We define a mirrored server set (a \emph{row}) as good if its instances respect the MZ balance implied by Constraint~\ref{mzconstraint:1}, and bad otherwise. The greedy algorithm iteratively repairs bad rows using swaps that restore them to good states while minimizing segment movement:

For each bad \textit{row} $S_{\mathrm{bad}}$ that contains one or more overpopulated MZs, select an instance from an overpopulated MZ and find another \textit{row} $S'$ (preferably another bad \textit{row}) such that, after swapping the two instances: (i) the number of overpopulated MZs in each of $S_{\mathrm{bad}}$ and $S'$ does not increase, and (ii) the total number of overpopulated MZs across $S_{\mathrm{bad}}$ and $S'$ decreases by at least one, ensuring that each step makes progress without worsening any row.

\textbf{Correctness Proof.}
We prove upscaling; node-swap/downscaling are analogous.

\noindent\textit{Base Case.}
For $R = 1$, each row has one instance, so no MZ repeats.

\noindent\textit{Inductive Hypothesis.}
Assume that for $R = r$, every row in the $r \times \mathrm{N_{i/rg}}$ assignment respects Constraint~\ref{mzconstraint:1}.

\noindent\textit{Inductive Step.}
Show the claim for $R = r + 1$.

\smallskip
\hspace{1em}\textit{Scenario 1: $r < \mathrm{MZ}$.}
Add one new instance $\delta$ to each row in column $r+1$, then run Algorithm~\ref{alg:swaptomzaware}.

Suppose, for contradiction, that a \emph{bad} row remains and no valid swap exists. Let there be $p \neq 0$ bad rows and $q$ good rows, where $p + q = \mathrm{N_{i/rg}}$.

Consider a bad row of the form
\begin{equation}
x_{i,1}, x_{i,2}, \dots, x_{i,r}, \delta_i \quad \text{for} \quad i \in \{1, \dots, p\},
\label{eq:bad-row}
\end{equation}
Since the row is bad and $r < \mathrm{MZ}$, $\delta_i$'s MZ already appears in $\{x_{i,1}, \dots, x_{i,r}\}$.

\begin{enumerate}
    \item Swapping with a good row.
    Any good row has the form
    \[
        x_{k,1}, x_{k,2}, \dots, x_{k,r}, \delta_k \quad \text{for} \quad k \in \{1, \dots, q\}.
    \]
    An invalid swap involving $\delta_i$ has two cases:
    \begin{enumerate}
        \item[(1a)] $\delta_i$ is already present in the good row, i.e.,
        \[
        \delta_i \in \{x_{k,1}, x_{k,2}, \dots, x_{k,r}, \delta_k\}.
        \]
        Inserting $\delta_i$ duplicates an MZ, so the swap is disallowed.

        \item[(1b)] $\delta_i$ is not present in the good row.
        Invalidity then requires every candidate from the good row to already appear in the bad row:
        \[
            \{x_{k,1}, x_{k,2}, \dots, x_{k,r}, \delta_k\} \subseteq \{x_{i,1}, x_{i,2}, \dots, x_{i,r}\}.
        \]
        This selects $r+1$ distinct elements from a size-$r$ set, impossible; thus Case (1b) cannot occur.
    \end{enumerate}
    Thus every good row must contain $\delta_i$.

    \item Swapping with another bad row.
    Any other bad row has the form
    \[
        x_{j,1}, x_{j,2}, \dots, x_{j,r}, \delta_j \quad \text{for} \quad j \neq i,\ j \in \{1, \dots, p\}.
    \]
    The swap is invalid only if both rows remain bad:
    \begin{align*}
        x_{i,1}, x_{i,2}, \dots, \delta_i, \dots, x_{i,r}, \delta_j &\quad \text{(bad)}, \\
        x_{j,1}, x_{j,2}, \dots, \delta_j, \dots, x_{j,r}, \delta_i &\quad \text{(bad)}.
    \end{align*}
    The second row remains bad only if $\delta_i$ appears in it:
    \[
        \delta_i \in \{x_{j,1}, x_{j,2}, \dots, x_{j,r}, \delta_j\}.
    \]
\end{enumerate}

Thus \emph{every} other row contains $\delta_i$.

Row $i$ has at least two instances of $\delta_i$'s MZ; the other $\mathrm{N_{i/rg}} - 1$ rows each have at least one. Thus the total is at least
$
    2 + (\mathrm{N_{i/rg}} - 1) = \mathrm{N_{i/rg}} + 1.
$
But when $r \le \mathrm{MZ}$, any single MZ can host at most
\[
    \frac{r \cdot \mathrm{N_{i/rg}}}{\mathrm{MZ}} \le \mathrm{N_{i/rg}}
\]
instances, making $\mathrm{N_{i/rg}} + 1$ impossible, a contradiction.

Hence the algorithm cannot get stuck; it makes progress until all rows satisfy the requirement.

\smallskip
\hspace{1em}\textit{Scenario 2: $r \geq \mathrm{MZ}$.}
Give each MZ $\lfloor r / \mathrm{MZ} \rfloor$ instances, then run Algorithm~\ref{alg:swaptomzaware} as above. The result has at most $\lceil R / \mathrm{MZ} \rceil$ instances from any MZ, and min/max MZ counts differ by at most $1$, satisfying Constraint~\ref{mzconstraint:1}.

Thus the claim holds for $R = r+1$ whenever it holds for $R = r$, completing the induction. Minimal movement follows because each accepted swap strictly reduces overpopulated MZs, repairs two rows when possible (at least one otherwise), and never reintroduces imbalance into fixed rows. Hence every swap is monotonic, and no valid state is reachable with fewer row updates.

\textbf{Validation.}
We validated the implementation with manually crafted adversary cases covering creation, replacement, uplift, downlift, reused-instance repair, drift, and multi-row swaps. We also ran an offline randomized correctness test over one million randomly generated imbalanced initial conditions, checking that repair produced complete assignments with no reused instances, and that mirrored server sets satisfied the MZ-diversity constraint.

\subsection{Impact-Free Data Rebalance}
\label{sec:impact-free-rebalance}

Given a target assignment---whether computed by the MZ-aware algorithm above or by any other placement policy---the system must transition from the current state to the desired state without degrading query performance. One option is to manage resource consumption during data loading (e.g., throttling downloads, limiting index builds). However, such a solution would introduce substantial complexity into segment-loading code for operations that are relatively infrequent.

We instead propose a simpler strategy: drain query traffic from hosts while they download substantial data. This approach is enabled by Pinot's data model, which is mostly immutable with multiple serving replicas per segment.

Pinot's architecture allows tables to scale in three dimensions:

\begin{enumerate}
    \item The replica count---scaling  horizontally with traffic
    \item The number of instances assigned to each replica---to accommodate changes in total data size or query patterns
    \item The host size---allocating more powerful hosts
\end{enumerate}

We aimed to develop a generic approach that can adjust any combination of these dimensions while satisfying the following safety conditions:

\begin{enumerate}
    \item \label{rebalance:safety1} Serving capacity reduction should be minimized during scaling. We should never have more than one unavailable replica. When increasing replication only, we should avoid any down replicas (if possible).
    \item \label{rebalance:safety2} At any stage in the algorithm, we should minimize the extra data hosted by any server, to avoid server disks filling or unbalanced segment load.
    \item \label{rebalance:safety3} Queries should be drained before substantial segments are added to a host. ``Substantial'' is defined as more segments than a regular data push for the table.
\end{enumerate}

\textbf{State Model.}
We formalize the rebalancing problem as a state-transition system. Let $H$ denote the set of all hosts and $S$ the set of all segments for a given table. At any point during rebalancing, we maintain the following per-host segment sets:

\begin{enumerate}
    \item $\mathit{Initial}(h)$: the segment assignment before rebalancing begins.
    \item $\mathit{Desired}(h)$: the target assignment after rebalancing completes (e.g., as computed by the MZ-aware algorithm in Section~\ref{sec:mz-aware}).
    \item $\mathit{Added}(h) = \mathit{Desired}(h) \setminus \mathit{Initial}(h)$: segments that must be downloaded.
    \item $\mathit{Removed}(h) = \mathit{Initial}(h) \setminus \mathit{Desired}(h)$: segments that must be dropped.
\end{enumerate}

At rebalance step $n$, the effective state of host $h$ is:
\[
    \mathit{Current}_n(h) = \mathit{Initial}(h) \cup \mathit{Added}_n(h) \setminus \mathit{Removed}_n(h)
\]
where $\mathit{Added}_n(h) \subseteq \mathit{Added}(h)$ and $\mathit{Removed}_n(h) \subseteq \mathit{Removed}(h)$ represent the segments added and removed so far. A host is \emph{converged} when $\mathit{Current}_n(h) = \mathit{Desired}(h)$. Additionally, we track $\mathit{Down}_n(h) \in \{0, 1\}$ to indicate whether a host has been drained of queries at step $n$. The rebalance terminates when all hosts are converged and serving.

\begin{algorithm}
\caption{Impact-Free Rebalance Algorithm}
\label{alg:impact_free_rebalance}
\begin{algorithmic}[1]
\Procedure{rebalance}{T}
    \State $goal \gets calculate\_desired(T)$
    \While{$goal \neq (current \gets current\_state(T))$}
        \State $goal \gets update\_if\_needed(goal)$
        \State enable\_converged\_hosts(current)
        \State $candidates \gets$ \Call{select\_hosts}{$current, goal$}
        \If{$candidates \neq \emptyset$}
            \ForAll{$host \in candidates$}
                \State disable\_queries($host$)
                \State wait\_for\_inflight\_queries($host$)
                \State assign\_segments(desired, host)
            \EndFor
        \Else
            \State $progress\_step(current, goal)$
        \EndIf
    \EndWhile
\EndProcedure

\Function{select\_hosts}{current, goal}
    \State $H \gets \emptyset$
    \For{$host \in by\_priority(hosts)$}
        \If{$\neg done(goal, host) \land safe(current, H, host)$}
            \State $H.add(host)$
        \EndIf
    \EndFor
    \State \Return $H$
\EndFunction
\end{algorithmic}
\end{algorithm}

\textbf{Algorithm Design.} The algorithm converges the current segment assignment to the desired state in a series of steps (Algorithm~\ref{alg:impact_free_rebalance}). In each step, one of two actions is taken:

\begin{enumerate}
    \item \emph{Rebalancing Step}: one or more hosts are fully rebalanced to their desired assignment after draining queries.
    \item \emph{Progress Step}: a small percentage of segments are added to all hosts when no hosts can currently be fully drained.
\end{enumerate}

A \emph{rebalancing step} proceeds as follows for each selected host $h$:
\begin{enumerate}
    \item Disable queries on $h$ (mark $\mathit{Down}_n(h) = 1$).
    \item Wait for all in-flight queries on $h$ to complete.
    \item Assign all remaining segments in $\mathit{Added}(h) \setminus \mathit{Added}_n(h)$ to $h$ and remove all segments in $\mathit{Removed}(h) \setminus \mathit{Removed}_n(h)$.
    \item Wait for segment downloads to complete on $h$.
    \item Re-enable queries on $h$ (mark $\mathit{Down}_{n+1}(h) = 0$).
\end{enumerate}

A host $h$ is considered \emph{safe} to rebalance at step $n$ if draining it would not reduce the available replicas for any segment below a configurable threshold. Formally, for every segment $s$ currently served by $h$:
\[
    \sum_{h' \in H \setminus \{h\}} \mathbf{1}[s \in \mathit{Current}_n(h') \land \mathit{Down}_n(h') = 0] \;\geq\; T
\]
where $T$ is the minimum required serving replicas (e.g., $R - 1$ for single-replica headroom). Our implementation prioritizes hosts with the largest difference between current and desired states, maximizing progress per step.

When no host can be safely picked---for example, because too many segments have reduced replication from prior steps---a \emph{progress step} is executed instead. This step adds a small number of segments (comparable to a table's regular daily data push) across all non-converged hosts, prioritizing segments with the fewest available replicas. Progress steps incrementally restore replication until hosts become safe to fully rebalance.

\textbf{Convergence and Safety.} The algorithm guarantees monotonic progress: each rebalancing step fully converges at least one host, and each progress step strictly increases the total number of assigned segments. Since the target state is
finite, termination is guaranteed. Safety Condition~\ref{rebalance:safety1} holds because the host-selection check ensures replica availability never falls below the required threshold. Safety
Condition~\ref{rebalance:safety2} holds because hosts converge in a single step—unneeded segments are removed immediately—and progress steps add only small increments. Safety Condition~\ref{rebalance:safety3}
holds because queries are drained before any substantial segment reassignment. Finally, the implementation is designed for fault tolerance and operational simplicity: if the operator process is terminated
mid-operation or the state store becomes unavailable, a subsequent invocation can resume from where it left off by deterministically recomputing the current state.


\subsection{Production Results}
\label{sec:rebalance-results}

We deployed both the MZ-aware assignment algorithm and the impact-free rebalance procedure in our production environment that spans 250+ clusters, 5{,}500+ hosts, and 4{,}500+ tables hosting $\sim$10~PB of data (Table~\ref{tab:deployment-scale}).

Over the past 12 months, MZ-aware assignment has sustained over 99.9\% production availability despite daily node churn of up to 7\% per cluster and more than 50 on-demand scaling events, with incremental
reassignments that avoid large-scale segment movement under steady-state operation.

During the past 24 months of serving, impact-free rebalance has guaranteed safety, idempotency, and minimum operational effort during cluster scalings---including an in-place migration of all Pinot production data to our MZ-aware layout within 6 months. This migration moved over 10~PB of segment data across 5{,}500+ hosts without a single SLA violation.

\subsection{Lessons Learned}

Our experience in production surfaced several complications:

\begin{enumerate}
    \item Pinot's assignment logic is deterministic given the current instance list and segment
    set, but is not stable across incremental rebalance steps; we therefore snapshot the desired
    assignment at the start and preserve it throughout convergence.
    \item Uncommitted (in-flight) segments could cause stale query results during a drain; we
    trigger a segment commit before disabling queries on a host to bound staleness.
    \item Tables with very high ingestion rates caused progress steps to stall because newly
    pushed segments continuously reset the convergence threshold; we addressed this with a
    small allowed deviation budget and an early-termination condition.
    \item Concurrent rebalance invocations can interfere with each other; we use
    ZooKeeper-based leader election so that at most one rebalance operator drives a given table
    at any time.
\end{enumerate}

\section{Adaptive Server Selection}
\label{sec:adss}

\subsection{Motivation}

The routing strategies described in Section~\ref{sec:adss-background} worked well initially, but organic traffic growth, onboarding of latency-sensitive applications, and cost-driven cluster consolidation required increasing the number of server instances per replica group. With replica-group-based routing, a single slow server---caused by background tasks, GC pressure, or a preceding expensive query---bounds the latency of every query routed to that group. Since the broker requires results from \emph{all} servers, the overall latency is dictated by the slowest server, degrading P95/P99 at the broker level.

\subsection{Solution}

ADSS adapts ideas from C3~\cite{suresh2015c3}, ELS~\cite{spotify_els_part2}, and Finagle Peak EWMA~\cite{finagle_peak_ewma} to Pinot's replica-group-based scatter-gather model using only broker-local observations, without server-side feedback or rate control. In Pinot's mirrored replica-group layout, servers in different replica groups may hold identical segment sets; we call each such equivalent group a \emph{Mirror Server Set} (MSS) (Figure~\ref{fig:routing}, right). Each broker independently scores servers from per-$\langle$table, server$\rangle$ in-flight counts and response latencies, then routes each query within every MSS toward lower-scored servers as conditions change.

We provide three scoring algorithms, configurable per table: (1)~\textit{In-flight requests}; (2)~\textit{Latency EMA}~\cite{wikipedia_moving_average}; and (3)~\textit{Hybrid}. The hybrid score adapts C3's queue-aware replica selection~\cite{suresh2015c3}, ELS's broker-local latency smoothing~\cite{spotify_els_part2}, and Finagle Peak EWMA's latency-times-load intuition~\cite{finagle_peak_ewma}:
\begin{align}
	\mathit{estimatedQ} &= \mathit{Request}_{\mathit{inflight}} + Q_{\mathit{EMA}} + 1 \\
	\mathit{Score} &= (\mathit{estimatedQ})^{N} \cdot \mathit{Latency}_{\mathit{EMA}}
	\label{eq:hybrid-score}
\end{align}
where $Q_{\mathit{EMA}}$ is the EMA of the broker's observed in-flight queue size, updated as $Q_{\mathit{EMA}} \leftarrow \alpha \cdot Q_{\mathit{current}} + (1{-}\alpha) \cdot Q_{\mathit{EMA}}$. The $Q_{\mathit{EMA}} + 1$ term forecasts near-future queue depth, $\alpha \sim 2/3$ balances responsiveness and stability, and the exponent $N$ is set near 3 following C3's cubic queue-penalty intuition~\cite{suresh2015c3}. To mitigate synchronized oscillation when multiple brokers independently select the same server, we optionally replace argmin selection with a softmax-based probabilistic policy. Given an MSS with servers $\{s_1, \dots, s_k\}$, each server is selected with probability:
\begin{equation}
	P(s_i) = \frac{e^{-\mathit{Score}(s_i)\,/\,\tau}}{\sum_{j=1}^{k} e^{-\mathit{Score}(s_j)\,/\,\tau}}
	\label{eq:softmax-select}
\end{equation}
where $\tau$ is a scaling factor derived from the score magnitudes to prevent overflow. This converts scores into selection probabilities---lower-scored (better) servers receive higher probability but not exclusively. The key benefit is \emph{oscillation control}: by introducing calibrated randomness, brokers no longer synchronize on the same ``best'' server at each scoring cycle, breaking the feedback loop that causes load swings between servers while still preserving effective traffic diversion from degraded servers (validated in Section~\ref{sec:adss-simulation}).

We choose $\tau$ from a target traffic-diversion property rather than as an absolute constant. The calibration is derived for a prolonged steady-state slowness scenario: one server is consistently slower than its peers, but no queue has built up yet. As a common example, if a workload normally operates at roughly 66\% of its SLA latency target, a 1.5$\times$ latency increase reaches the SLA boundary. We therefore treat a server whose score is roughly 1.5$\times$ the healthy-server score as slow enough to receive negligible traffic.

Consider an MSS with $K$ servers, where $K{-}1$ healthy servers have score $S$ and one slow server has score $rS$. Under Eq.~\ref{eq:softmax-select}, the slow server receives probability:
\begin{equation}
P_{\mathit{slow}} = \frac{1}{(K{-}1)e^{(r-1)S/\tau}+1}
\label{eq:tau-calibration}
\end{equation}
Requiring $P_{\mathit{slow}} < 0.1\%$ gives $\tau < (r{-}1)S / \ln(999/(K{-}1))$. For a typical 3-server MSS and $r{=}1.5$, this yields $\tau \lesssim 0.08S$. Since brokers do not know the healthy-server score in advance, we estimate $S$ from the current average score in the MSS; for $K{=}3$, this corresponds to $\tau \approx 0.07$ times the current average score. For MSS sizes from 3 to 10, the same calibration is generally applicable.

The same $\tau$ selection also covers transient slowness. When a server stalls briefly, its in-flight queue builds up quickly, and the cubic queue penalty in Eq.~\ref{eq:hybrid-score} drives its score well beyond 1.5$\times$ the healthy score. Thus the softmax policy diverts traffic for both prolonged slowdowns and short-lived latency spikes across the workloads we evaluated.

\subsection{Numerical Simulation}
\label{sec:adss-simulation}

To validate parameters and behaviors difficult to test in production, we built a discrete-time simulator of Pinot's broker-server architecture. Time advances in 0.1\,ms increments; queries arrive according to the configured workload, brokers dispatch through ADSS, servers process segment tasks, and completions update broker EMA statistics. Server degradation is probabilistic: a slow server's threads make progress with probability $p < 1$ (e.g., $p{=}0.4$ for 60\% throughput reduction), matching latency variation from GC pauses and I/O contention. Identical arrival sequences isolate selector behavior across base latency, QPS, broker count $B$, replicas $S$, EMA smoothing $\alpha$, exponent $N$, and latency prior.

\begin{figure}[htbp]
  \centering
  \includegraphics[width=0.95\columnwidth]{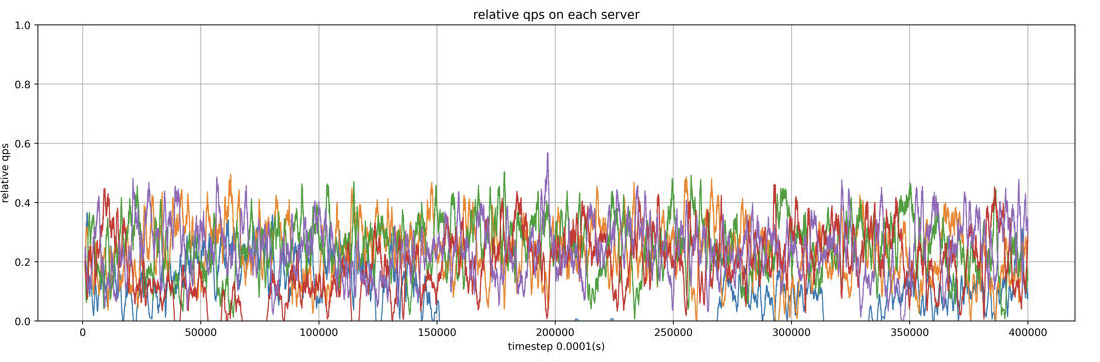}
  \vspace{2pt}
  \includegraphics[width=0.95\columnwidth]{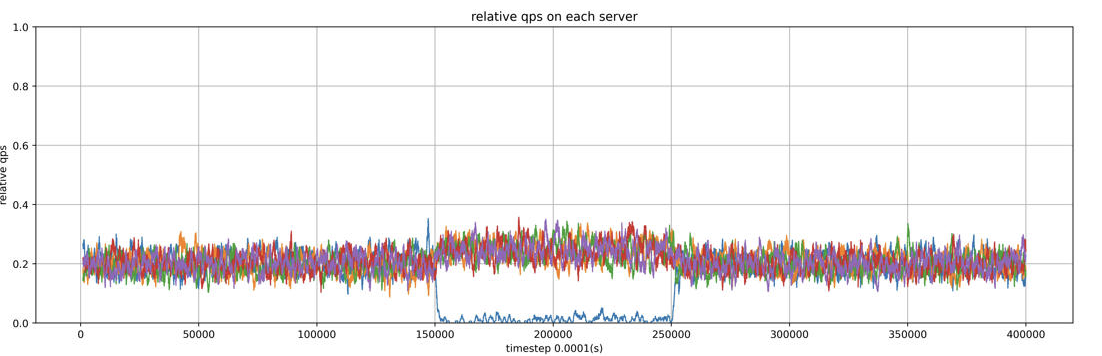}
  \caption{Relative QPS per server over time: hybrid selector (above) vs.\ softmax variant (below).}
  \label{fig:adss-oscillation}
\end{figure}

\textbf{Strategy Comparison.}
We swept four workload profiles---high QPS / low latency (1500\,QPS, 1.35\,ms), moderate QPS / moderate latency (300\,QPS, 20\,ms), low QPS / high latency (30\,QPS, 200\,ms), and very high QPS / low latency (6000\,QPS, 1\,ms)---as well as a hybrid workload combining all three tiers (2400\,QPS at 1\,ms $+$ 200\,QPS at 10\,ms $+$ 40\,QPS at 100\,ms). Table~\ref{tab:adss-sim} summarizes the behavioral differences across the four selectors.

\begin{table}[h]
\centering
\caption{Simulated selector behavior with one degraded server.}
\label{tab:adss-sim}
\small
\begin{tabular}{lccc}
\toprule
\textbf{Selector} & \textbf{Diversion} & \textbf{Oscillation} & \textbf{Recovery} \\
\midrule
Round Robin   & None     & N/A  & N/A \\
In-Flight     & Fast     & High & Immediate \\
Latency EMA   & Gradual  & Low  & Slow \\
Hybrid        & Fast     & Low  & 2--3\,s \\
\bottomrule
\end{tabular}
\end{table}

Across all profiles, the hybrid selector diverted traffic from the degraded server within 2--3 scoring cycles while maintaining stable QPS on healthy servers. The in-flight selector reacted equally fast but oscillated under high QPS due to lack of memory of past slowness. The latency-EMA selector avoided oscillation but adapted slowly under low-QPS workloads. On the hybrid workload, the hybrid selector generalized across all three latency tiers with a single parameter configuration.

\textbf{Softmax-Based Oscillation Control.}
We evaluated the softmax policy (Eq.~\ref{eq:softmax-select}) on configurations with the most visible oscillation (1500QPS, 3 brokers, 5 replicas). Figure~\ref{fig:adss-oscillation} compares the deterministic hybrid selector with the softmax variant: the softmax variant substantially smoothed QPS distribution across servers while preserving traffic diversion from degraded servers, confirming its effectiveness for oscillation control.

\textbf{Edge Cases.}
The simulation has also identified a few edge cases that would have caused production incidents without proper patching. (1)
\emph{Cold-start with a slow server}: if a server is degraded from startup, $\mathit{Latency}_{\mathit{EMA}}$ stays at zero, causing the hybrid score to funnel queries to the bad server. A small latency prior (e.g., 1\,ms) eliminates this. (2) \emph{Low-headroom cascading}: when all servers operate near capacity and one degrades, redirected traffic overwhelms healthy servers. This motivates server headroom resize and server-side queue-depth limits---both deployed in production.

\subsection{Production Evaluation}
\label{sec:adss-evaluation}

We evaluated Adaptive Server Selection in production at LinkedIn, where Pinot serves over 250K queries per second across 4,500+ server hosts, 1,000+ broker hosts, and 4,500+ tables (Table~\ref{tab:deployment-scale}).

\textbf{Routing Overhead.}
We benchmarked the hybrid selector against round-robin at 1,500 QPS using production query traces. Table~\ref{tab:adss-overhead} reports the results.

\begin{table}[hbpt]
\centering
\caption{Routing-phase and end-to-end query latency (ms) at 1,500 QPS.}
\label{tab:adss-overhead}
\small
\begin{tabular}{lcccccc}
\toprule
& \multicolumn{3}{c}{\textbf{Routing Phase}} & \multicolumn{3}{c}{\textbf{Total Query Processing}} \\
\cmidrule(lr){2-4} \cmidrule(lr){5-7}
& P50 & P95 & P98 & P50 & P95 & P98 \\
\midrule
Baseline  & 0.014 & 0.021 & 0.024 & 4.6 & 5.6 & 6.5 \\
Adaptive  & 0.019 & 0.030 & 0.033 & 4.6 & 5.4 & 5.9 \\
\bottomrule
\end{tabular}
\end{table}

The routing-phase overhead is sub-millisecond (under 0.01\,ms increase at P50), negligible relative to total processing times of 4--7\,ms. At P98, total query latency \emph{improved} by approximately 9\%, as the adaptive selector avoids routing to slower servers even under normal conditions.

\textbf{Effectiveness Under Server Slowness.}
We evaluated two failure modes in production. For \emph{prolonged slowness} (disk failure, persistent GC pressure), the hybrid selector routes traffic away within seconds, maintaining stable P95/P99 broker latencies. For \emph{transient slowness} (GC pauses, compaction), the exponential penalty ($N{=}3$) shifts traffic within one scoring cycle; as the server recovers, the score drops sharply, restoring traffic in 2--3\,s. Transient spikes that previously caused P90 latency to exceed 1.5$\times$ baseline were contained to under 10\% deviation.

\begin{figure}[hbt]
  \centering
  \includegraphics[width=0.95\columnwidth]{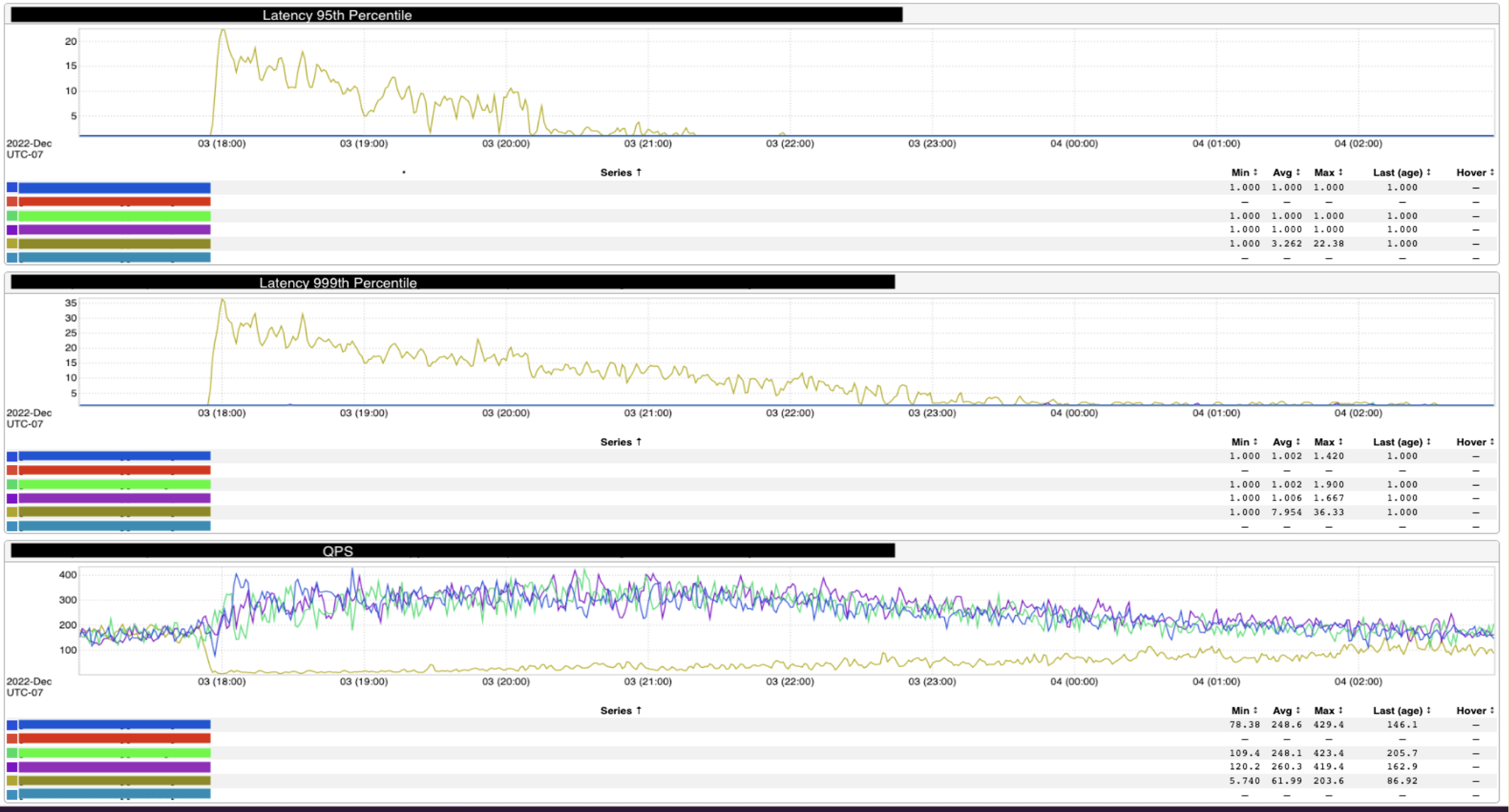}
  \caption{ADSS in JVM misconfiguration incident: P95/P999 latency (top two) and QPS distribution (bottom).}
  \label{fig:adss-production}
\end{figure}

Figure~\ref{fig:adss-production} illustrates a representative production incident. A misconfiguration in the JVM heap size on one Pinot server caused elevated query latencies ($\sim$20\,ms at P95, versus sub-millisecond on healthy servers). The adaptive selector detected the degradation and routed QPS away (yellow) from the affected server to near zero within seconds. Once the issue was gradually resolved, QPS steadily ramped back up to the recovering server without manual intervention.

\textbf{Latency Degradation Prevention.}
With replica-group routing and a replication factor of 3, a single slow server causes latency degradation for approximately 33\% of queries. Adaptive Server Selection breaks this coupling by selecting within each MSS independently. Over a 12-week production window (January--March 2023), the \emph{latency degradation prevention rate} consistently exceeded 90\%, meaning fewer than 10\% of queries experienced degradation during slowness events, compared to $\sim$33\% under the baseline.

\textbf{Operational Benefit.}
Slow-server alerts decreased by 97\% (60$\to$2 per quarter) and engineering hours on slowness investigations by 89\% (72$\to$8\,h per quarter). Residual alerts correspond to cluster-wide events where \emph{all} servers in an MSS degrade simultaneously---motivating the workload isolation mechanisms in Section~\ref{sec:QWI}.

\section{Related Work}
Our work addresses four interrelated challenges in operating large-scale OLAP systems. We briefly survey related work and highlight key differences.
\label{sec:related}

\subsection{Workload Management and Resource Isolation}
Classical and commercial warehouses expose workload managers via admission control, scheduling, and static resource groups~\cite{niu2011,tozer2012,ibm-db2-wlm,oracle-rm,saphana-wlm}. Cloud warehouses---BigQuery~\cite{bigquery-slots}, Redshift~\cite{redshift-wlm}, Snowflake~\cite{dageville2016snowflake}---manage resources at the cluster or pool level through slots or virtual warehouses. Query-level scheduling has been explored in WiSeDB~\cite{marcus2016wisedb}, morsel-driven parallelism~\cite{leis2014morsel}, and IconqSched~\cite{iconqsched2023}, and multi-tenant isolation in F1 Query~\cite{samwel2018f1}, Liminal~\cite{shukla2023liminal}, and SQLVM~\cite{narasayya2013sqlvm}. Cluster schedulers such as YARN~\cite{yarn-capacity}, Kubernetes~\cite{k8s-resourcequota}, and Borg~\cite{verma2015borg} enforce fairness at the job or container level.
 
Closer to QWI's design space, Apache Doris~\cite{apache-doris-wlg} 
and StarRocks~\cite{starrocks-resource-group} delegate per-workload 
CPU and memory enforcement to Linux cgroups~\cite{linux-cgroups-v2}. 
TiDB's Resource Control~\cite{tidb-resource-control} is more closely 
related to QWI: it abstracts CPU and IOPS into Request Units (RUs) 
enforced via a token-bucket algorithm, which is structurally similar 
to QWI's per-window allowance model. The algorithmic contrast lies 
in \emph{what} is charged and \emph{when}: RU-based schemes 
pre-classify each operation into a synthetic composite cost 
(combining CPU time, IOPS, and RPC bytes) computed at admission, 
while QWI charges raw, measured CPU-ns and allocated bytes sampled 
during execution. Pre-classification works well for OLTP where 
per-operation costs are bounded and predictable, but mis-estimates 
in the OLAP regime where the same operator can vary in cost by 
orders of magnitude depending on data distribution and selectivity. 
The cgroup-based approaches face a separate barrier: Pinot servers 
run as one JVM hosting many concurrent queries across shared thread 
pools, so cgroup boundaries cannot attribute usage to workloads 
sharing a JVM, and one container per workload would shard resources, defeating the multiplexing benefit of shared cpu/jvm heap/mmap buffer and inflate operational 
cost at LinkedIn's scale.

QWI occupies a middle ground for in-engine, multi-tenant JVM deployments: it treats CPU time and allocated bytes as continuously enforced per-workload budgets via sampling-based accounting that requires no operator changes, and enforces them inside both brokers and servers with sub-millisecond responsiveness and negligible overhead. Higher-level schedulers can consume QWI's cost signals while relying on QWI for fine-grained enforcement.
\subsection{Adaptive Query Routing}

The ``tail at scale'' problem~\cite{dean2013tail} shows how latency variability compounds under fan-out. C3~\cite{suresh2015c3} combines queue size estimation with latency measurements for Cassandra, achieving up to 3$\times$ improvement at P99.9. H\'eron~\cite{heron2021} extends replica selection to heterogeneous workloads, and Spotify's ELS~\cite{spotify_els_part2} uses exponential decay for latency estimation. ClickHouse Cloud's parallel replicas~\cite{clickhouse_parallel_replicas} use a pull-based execution model: replicas request scan tasks as they make progress, faster replicas can steal remaining work, and unavailable replicas can be excluded from the query. Recent fail-slow work studies broader system-level adaptation: ADR~\cite{lu2025onesizefitsnone} adapts detection thresholds using recent value and update-frequency histories before invoking mitigation actions, while Odyssey~\cite{zhou2025odyssey} selects among recovery strategies for resilient distributed training. Database-specific approaches include Cassandra's dynamic snitch~\cite{cassandra-snitch}, MongoDB's nearest member routing~\cite{mongodb-readpref}, and CockroachDB's~\cite{taft2020cockroachdb} follow-the-workload. Service mesh load balancers like Envoy~\cite{envoy-lb} and L3~\cite{michaelis2024l3} operate at the service level.

Our approach differs by exploiting Pinot's mirrored replica group layout: rather than treating replicas independently or implementing a general fail-slow recovery framework, ADSS selects the best-scoring server within each Mirror Server Set (MSS) during broker-side scatter-gather routing. ClickHouse's pull-based task scheduling is not directly aligned with Pinot's replica-group-based scatter-gather execution, where a broker must choose servers before dispatching each query. ADSS nevertheless provides analogous mitigation at the routing layer: slow servers receive less traffic, fast servers receive more traffic, and sufficiently degraded servers can be avoided by the routing decision itself. Unlike ADR's detect-then-mitigate model or Odyssey's post-failure recovery-policy selection, ADSS uses the same threshold-free scoring heuristic for lightweight slowness inference and mitigation. This preserves the operational benefits of symmetric assignment while improving query latency resilience using Pinot-specific routing signals.

\subsection{Online Data Rebalancing}

Zero-downtime migration has been addressed by Shopify's Ghostferry~\cite{shopify-ghostferry} for MySQL, Facebook's TAO~\cite{bronson2013tao} via dual-write cutover, and Google Spanner~\cite{corbett2013spanner} through automated load balancing. Systems like Vitess~\cite{vitess2015}, CockroachDB~\cite{taft2020cockroachdb}, and TiDB~\cite{huang2020tidb} provide automatic rebalancing for transactional workloads, while Druid~\cite{yang2014druid} coordinates segment movement through its coordinator.

These approaches focus on data consistency but often overlook query latency impact. Our algorithm explicitly drains queries before substantial segment movement, uses a two-phase approach (rebalancing steps for safe hosts, progress steps when blocked), and respects replica availability thresholds throughout.

\subsection{Fault Domain and Zone Awareness}\label{sec:relatedFd}

CRUSH~\cite{weil2006crush} provides topology-aware placement for Ceph; CRUSHED improves uniformity in Apache Helix~\cite{LiCRUSHEDBlog,ApacheHelixCRUSHED}. Cassandra's NetworkTopologyStrategy~\cite{cassandra-topology}, YugabyteDB~\cite{yugabyte-zone}, and CockroachDB~\cite{cockroach-zones} support zone-aware replication. Kubernetes provides topology spread constraints~\cite{k8s_topology_spread} and pod disruption budgets~\cite{k8s_pdb} for zone-level fault tolerance.

The built-in CRUSHED algorithm is appealing for its stable, even distribution with weight-based assignment. However, our evaluation surfaced four issues specific to our migration and Pinot deployment. First, migrating from our existing segment assignment to CRUSHED's consistent hashing mapping produces a large data shuffle, adding both operational risk and overhead at scale. We were able to reduce this movement by seeding the algorithm with the existing assignment — particularly important given the mismatch between our typical replica count (3) and the number of maintenance zones (10–20). Second, we rely heavily on strict replica-group routing to minimize query fan-out for latency-sensitive tables; this constraint is not natively incorporated into CRUSHED and required custom integration. Third, CRUSHED does not minimize data movement when replacing or adding server instances. Our mirror-server model enables 1:1 server replacements and replica additions with zero data movement in many cases — an important property we could not guarantee when replacement nodes may come from a different maintenance zone. Fourth, mirror servers simplify adaptive server selection: because mirrored instances host identical data, comparing "fast" versus "slow" servers within a mirror set is straightforward and does not require cross-segment reasoning.

\section{Conclusion}
\label{sec:conclusion}

This paper presented a holistic resiliency framework for Apache Pinot that unifies workload, structural, and runtime failure vectors into a single layered defense. Our experience at LinkedIn demonstrates that no single mechanism is sufficient for petabyte-scale OLAP resilience; rather, each layer must cover the failure modes left open by the others. Workload Resilience (QWI) bounds logical "noisy-neighbor" interference within shared clusters, while Runtime Resilience (ADSS) mitigates transient "fail-slow" behavior that logical isolation cannot detect. These mechanisms rely on the physical stability provided by Structural Resilience (Zone-Aware Placement and Impact-Free Rebalancing), which ensures data availability across fault domains during both correlated failures and large-scale data migrations.

Several limitations remain to be addressed. Currently, QWI assumes separately calibrated budgets across heterogeneous hardware and does not yet support automatic hardware-normalized accounting or quota borrowing. ADSS improves routing under localized slowness, but cannot dynamically increase cluster capacity during systemic failures. Furthermore, Zone-Aware Placement assumes roughly balanced node availability across maintenance zones. These limitations point to future work on heterogeneous-aware budgeting, coordinated interaction between workload isolation and adaptive routing, and richer control-plane cost modeling. 

Overall, our results show that sustaining low-latency OLAP at production scale requires coordinated controls across resource accounting, topology-aware placement, and adaptive routing. The mechanisms described here have enabled Pinot at LinkedIn to maintain 99.9\% availability and stable query latency SLAs while managing over 10 PB of data across 5,500+ hosts. This framework provides a production-proven architecture for a resilient, mission-critical online analytics infrastructure in the face of continuous operational churn. 


\end{document}